\documentclass[twocolumn,reprint,floatfix,prb,a4paper]{revtex4-2}

\usepackage{color}
\usepackage{graphicx}
\usepackage{amssymb}
\usepackage{amsmath}
\usepackage{hyperref}
\usepackage{bm}

\newcommand{\en}{\varepsilon}

\begin{document}

\title{Solvable coupled cluster states in $d=\infty$ with prescribed momentum distribution}
\author{Izak Snyman}
\affiliation{Mandelstam Institute for Theoretical Physics, School of Physics, University of the Witwatersrand,
Johannesburg, South Africa}
\author{Serge Florens}
\affiliation{Univ. Grenoble Alpes, CNRS, Grenoble INP, Institut Néel, 38000 Grenoble, France}

\begin{abstract}
Two of the most successful ideas in many-body theory, the exponential coupled cluster ansatz of
quantum chemistry and the dynamical mean-field theory (DMFT) in condensed matter physics, have 
to date, by-and-large developed separately. We establish a connection between them, by showing that 
the infinite dimensional limit of the hypercubic lattice allows the exact closed-form evaluation of static expectation 
values of coupled cluster wave functions, provided the cluster amplitudes admit a low-rank 
decomposition.
As a result, evaluation errors -- a serious concern in conventional coupled cluster theory -- 
can be disentangled from limitations to expressivity of coupled cluster states.
For this family of coupled cluster states, the cavity construction in $d\to\infty$ exactly reduces 
the environment of a lattice site to a Gaussian Grassmann kernel, 
and irreducible diagrams for one-body expectation values are found to be site-diagonal (in analogy 
to the DMFT self-energy).
The resulting ``impurity problem" is an integral over a finite number of Grassmann variables rather
than a quantum impurity problem in imaginary time, so that static observables derive exactly from 
a static self-consistency condition: a wave function counterpart of DMFT.
For the simplest case of a cluster amplitude restricted to rank-one doubles, we obtain the momentum 
distribution in a closed form. We show that any particle-hole symmetric momentum distribution
can be realized by a state that is the exact and unique ground state of an explicit parent Hamiltonian 
with a gapless single-particle tunnelling spectrum.
By construction these coupled cluster states variationally recover low order perturbation theory for the
Hubbard model. Their failure in the strong coupling regime near the Mott transition diagnoses a
fundamental limitation of coupled cluster doubles.
Intriguingly, our family of coupled cluster wave functions includes a gapless state 
with zero quasiparticle weight, possessing a momentum distribution that is completely smooth at the Fermi energy.\end{abstract}

\maketitle

\section{Introduction}

Correlated trial states are an invaluable resource to study interacting many-electron problems, from
the Jastrow and Gutzwiller wave functions to Slater-Jastrow states with backflow correlations
\cite{Jastrow1955,Gutzwiller1963,Gutzwiller1965,Feynman1956,Kwon1993,Foulkes2001,Becca2017}. There
is however an extreme tension between freedom in constructing them and the ability to calculate
observables in the assumed state. Some well-established methods resort to strategies such as
uncontrolled approximations, truncation of infinite hierarchies of equations at a low
order~\cite{BartlettMusial2007}, or Monte Carlo evaluations that are subject to sampling
noise~\cite{Ceperley1977,Foulkes2001,Becca2017}. It is often hard to disentangle limitations of a
trial state from errors introduced in evaluating it. Consequently it is desirable to have trial
states that can be evaluated exactly, yet are general enough that one can adjust targeted
observables freely. Importantly, such states may be useful independently of whether they produce
competitive variational energies. Their utility is to provide quantitative understanding of how a
particular state structure controls a particular observable and to allow rigorous benchmarking of
approximate evaluation procedures.

In this Article we ask whether it is possible to freely choose the single particle momentum
distribution $n_{\bm k}$ of a pure, translationally invariant, number-conserving, correlated,
many-electron state while retaining the ability to calculate local many-body correlations. The
question emerged out of an effort to complement the picture provided by the Gutzwiller trial
state~\cite{Gutzwiller1963,Gutzwiller1965}, which is designed to control double-occupation of
individual sites by lattice fermions, and which can be evaluated inexpensively and exactly on a
hypercubic lattice in the limit of dimension $d\to\infty$
\cite{Metzner1989,Metzner1989b,Muller-Hartmann1989,Gebhard1990}. It successfully captures
qualitative features of the Mott transition from a metal to an insulating state at a finite 
value of the Hubbard repulsion $U$~\cite{Hubbard1963,Brinkman1970}. Yet in the regime where local charge 
fluctuations are only moderately suppressed, the Gutzwiller state does not describe the correlated Fermi 
liquid particularly well: it possesses a momentum distribution that is piece-wise constant, taking on a
value $(1\mp Z)/2$ above or below the Fermi energy~\cite{Vollhardt1984,Metzner1989b}, with $Z$
the quasiparticle weight. It is thus of interest to build a correlated trial state that can be evaluated 
exactly and as simply as the Gutzwiller state in infinite dimension, but that better captures the single-particle 
momentum distribution of a correlated Fermi liquid. 

Our main methodological advance shows that a large class of coupled cluster wave functions
is amenable to exact, truncation-free computations in the limit of large dimension.  Coupled cluster
states are generally built from the exponential of a cluster operator, whose argument contains only 
the 
fermion creation and annihilation operators that do not annihilate a given reference 
(here the free Fermi sea). 
We consider coupled cluster states for lattice electrons.
In the limit of large dimensions, we show that the environment of a selected site enters through a self-consistent 
quadratic kernel. The construction borrows the cavity and embedding arguments underlying 
the dynamical mean-field theory (DMFT)~\cite{Georges1996}, 
that maps an interacting lattice problem onto a self-consistently determined interacting impurity problem 
in a non-interacting environment. However, the objects in our method are static expectation values 
rather than the frequency-dependent Green functions calculated with DMFT. The ``impurity problem" to 
be solved with our method only consists of integrals over a finite number of Grassmann variables associated 
with a single site, as opposed to path integrals over Grassmann fields of imaginary time. An answer is 
obtained either by elementary integration or by mapping back to fermions, and calculating static expectation 
values of a few-body problem. We stress that our main achievement is the exact computation of
coupled cluster wave functions in the $d\to\infty$ limit, and their use as variational states, using
a fruitful analogy borrowed from DMFT. This is in contrast to previous works~\cite{ChanCC,ZgidCC}, which utilized 
coupled cluster theory as solvers for the dynamical mean-field theory or embedding methods.

Coupled cluster theory with singles, doubles and perturbative triples is widely regarded as the 
reference method for weakly correlated molecules~\cite{Coester1958,Raghavachari1989,BartlettMusial2007}. 
Coupled cluster states are known to fully contain many-body perturbation theory (at a given order), 
while resumming entire classes of processes to all orders, making it in principle non-perturbative.
However, in contrast to our work, standard coupled cluster theory does not employ exact evaluation 
of expectation values. Rather a projection method is used that leads to a hierarchy of equations that 
would produce an exact result if evaluated to all orders, but that are only feasible to evaluate to 
up to low order. For instance, truncation at order $n=2$ (doubles)~\cite{Cizek1966,Purvis1982}
takes into account up to two-body terms in the argument of the exponent. Errors in evaluation are expected 
to be of the same order as those made when projecting the true ground state onto the doubles family. 
The advantage of this truncation scheme over schemes such as Configuration Interaction is that it produces 
size-consistent states~\cite{BartlettMusial2007,Shavitt2009}. Owing to the exponentiation of particle-hole creation
operators, a coupled cluster doubles state contains a number of excitations that is proportional to
the size of the system, rather than to the order $n=2$ of truncation. On the downside, the method
breaks down when correlations are strong~\cite{BartlettMusial2007}. When applied to 1-dimensional
Hubbard chains, truncated equations become singular at large Hubbard $U$
\cite{Paldus1984,Stein2014,Bulik2015,Qiu2017}. For large systems, it has proven difficult to
disentangle the errors incurred in the evaluation method (truncated hierarchy) from limitations of
the state (coupled cluster doubles) \cite{Wahlen-Strothman2018,Ai2026} because an exact variational
energy calculation is unfeasible.

The present work uncovers a setting where an exact evaluation of a family of coupled cluster
states can be performed in the thermodynamic limit, by using simplifications brought about by the limit of high
dimensions. (A recent study by the two of us~\cite{1dCoupledCluster} 
has shown that a class of coupled cluster states for charge insulators could also be tackled in the
thermodynamic limit in the opposite setting of one dimension, using exact algebraic structures related to matrix product states.)
The key building blocks of these cluster states are linear combinations of creation or annihilation 
operators of momentum orbitals in which the amplitude for momentum $\bm k$ is proportional to an energy envelope
$F_\alpha(\en_{\bm k})$. We allow distinct envelopes $\alpha=1,2,\ldots$, referred to as flavors. 
In fact, these flavors corresponds to the orbitals obtained from a low-rank Tucker decomposition of 
the cluster operator~\cite{TuckerCoupledCluster}.
By choosing envelopes that are small far from the Fermi energy, we can suppress high-energy particle-hole 
excitations, in a manner similar to the denominators of the perturbation expansion in $U$. 
Apart from its relevance in condensed matter, our work thus contributes to efforts at understanding 
limitations of coupled cluster theory.
It should be noted though, that our work does not allow for exact evaluation of expectation values 
when (after imposing momentum conservation) coupled cluster states further depend on single-particle momenta other than through $\en_{\bm k}$. 
A completely general form of the cluster operators would necessitate further methodological
extensions, beyond utilizing the collapse of irreducible self-energy-like diagrams to a site-local form in infinite dimension, 
that the current work relies on.

For simplicity, we limit our study to paramagnetic half-filled states (particle-hole symmetric) on
the hypercubic lattice. We find that any single-particle momentum distribution of the form 
$n_{\bm k}=n(\en_{\bm k})$ with
$n(\en)=1-n(-\en)$ can be realized by the simplest member of our trial state family, namely the
single-flavor doubles ansatz, and that there is a simple relation between the energy envelope and 
occupation probability:
\begin{equation}
n(\en)=\frac{F(\en)^2}{\mathcal{G}_0^{-3}+F(\en)^2}\leftrightarrow
F(\en)^2=\mathcal{G}_0^{-3}\frac{n(\en)}{1-n(\en)}
\end{equation} 
for $\en>0$, where 
$\mathcal{G}_0$ is the static analog to the DMFT Weiss field.
For a translationally
invariant, number-conserving, pure state, the amount by which $n(\en)$ differs from the step
function $\theta(-\en)$ is a measure of how far the trial state is from a single Slater determinant.
Our construction thus allows for the exact evaluation of local expectation values of states
arbitrarily far from an uncorrelated Fermi sea. Remarkably, we can construct not only states with
momentum distributions that have a jump at the Fermi energy $\en=0$, but also ones with a completely
smooth distribution. What is more, the gossamer metal construction \cite{Kollar2003} allows us to
write down an ungapped correlated hopping parent Hamiltonian that has any given trial state,
including choices with smooth $n(\en)$, as its unique ground state.


The main precedents for our theory are the Momentum-dependent Local Ansatz (MLA) \cite{Kakeshi2008,Patoary2013}, 
the Ghost Gutzwiller Ansatz (GGA) \cite{Lanata2017} and generalizations~\cite{Carlos2026}, 
and Variational Discrete Action Theory (VDAT) \cite{Cheng2021,Cheng2021b}, 
each being an embedding theory that maps the Hubbard model onto a single-site problem that is more straight-forward to solve than the full dynamical impurity problem encountered in DMFT. 
MLA also targets control over the single-particle momentum distribution. It builds on the local ansatz of Stollhoff and Fulde~\cite{Stollhoff1980}, in which the exponential of local correlation operators acts on a mean-field reference state.
 It however does not attempt to evaluate expectation values exactly. Instead it opts for the simple but uncontrolled approximation of neglecting correlations on sites other than the one considered. The GGA enriches the Gutzwiller ansatz by introducing auxiliary degrees of freedom that are eventually projected out. In contrast, our ``flavors'' refer to linear combinations of the actual fermionic degrees of freedom on the lattice. Our construction is designed to yield a Grassmann number problem, rather than the more nontrivial one involving Grassmann fields of imaginary time, that DMFT produces. VDAT can be seen as an intermediate option, that replaces the full imaginary time line with a few discrete time points. Evaluation of expectation values is exact in infinite dimension. For a coarse discretization, the complexity is comparable to that of our method. In contrast to VDAT, our construction employs two-body operators that are not fully localized to a single lattice site. This gives our construction more direct control of the single-particle momentum distribution than VDAT or GGA. There is also a body of work
\cite{Otsuka1992,Eichenberger2007,Eichenberger2009,Baeriswyl2009} initiated more than
three decades ago, aimed at improving the Gutzwiller ansatz by acting on it with the exponential of the kinetic energy operator. Originally these states were evaluated by brute force on small systems or by Monte Carlo methods. VDAT can be seen both as generalization of this work, and an exact evaluation of some of these states in infinite dimension. 

We consider the half-filled Hubbard model to assess the correlations represented by our family of
coupled-cluster states, restricting attention to states without magnetic order.  We stress that here
it is not our goal to offer a competitive ground state solver for the Hubbard model.    At weak
coupling, optimization of a single envelope recovers $98.6\%$ of the exact second-order in $U$
correlation-energy, compared with $84\%$ for the Gutzwiller state, and $66\%$ of the second-order
contribution to the jump in $n_{\bm k}$ at the Fermi energy, compared with $22\%$ for the Gutzwiller
state. The strong-coupling behavior is more restrictive: the optimal single-flavor state retains a
double occupancy of approximately $0.1283$, rather than approaching complete suppression of local
charge fluctuations, that the Gutzwiller ansatz can reach. These results distinguish the ability to
reproduce a detailed momentum distribution from the ability to describe the local correlations
required at strong coupling. 

The rest of the Article is structured as follows. In Sec.~\ref{sec:construction} we give the
details of our construction.  In Sec.~\ref{sec:eval} we obtain our main methodological result --
the exact evaluation of expectation values of arbitrary observables involving fermions associated
with a single site, as well as all single-particle additive operators. In Sec.~\ref{sec:givendelta}
we present closed form results for the simplest state in the family we consider, namely the
single-flavor ansatz. We calculate the energy envelope that produces a given particle-hole symmetric
single-particle momentum distribution, and obtain a parent Hamiltonian. In Sec.~\ref{sec:Hubbard}
we calculate the variational energy for the Hubbard model with respect to the single-flavor ansatz.
In Sec.~\ref{sec:beyonddoubles} we add another flavor, as well as the triples and quadruples
cluster operators allowed by the second flavor. A brief summary and conclusions are provided in
Sec.~\ref{sec:conclusions}. Various technical details are provided in four appendices. 
Appendix \ref{app:A} proposes an alternative to the usual derivation of the expression for the quadratic 
environmental action in terms of the local Green function and self-energy at the heart of the self-consistent
cavity method. This derivation applies both to the evaluation of static wave functions and 
to dynamical Green functions.
The generalization of the formalism required to include multiple flavors and higher particle-hole 
excitations (including triples and quadruples in the cluster operator) is covered in 
Appendices~\ref{app:doubocc2}, \ref{app:C}, \ref{app:3and4}.

\section{General construction}
\label{sec:construction}
\subsection{Design principles}
We present in this section the rationale leading to the design of coupled cluster states
that are amenable to the $d\to\infty$ limit. While we will have mostly the Hubbard model in mind, 
no reference to a specific model Hamiltonian needs to be made at this stage.
We take as vacuum a Fermi sea $\left|F\right>$ of spin-$1/2$ non-interacting fermions on a
$d$-dimensional hypercubic lattice with nearest neighbor hopping amplitude $-t^*/\sqrt{d}$,
and all other hopping amplitudes zero. The orbital with Bloch momentum $\bm k$ has energy
\begin{equation}
\en_{\bm k}=-\frac{2t^*}{\sqrt{d}}\sum_{a=1}^d\cos k_a.\label{eq:dispersion}
\end{equation}
All Bloch momentum orbitals with $\en_{\bm k}<0$ are filled (the Fermi energy is set to zero
throughout, and we stick to particle-hole symmetry).
When local interactions are introduced, the scaling $1/\sqrt{d}$ of hopping amplitudes produces 
a non-trivial ground state in the $d\to\infty$ 
limit~\cite{Metzner1989b,Muller-Hartmann1989}, while making the many-body problem more tractable
using DMFT~\cite{Georges1996}.

A paradigmatic many-body wave function that can be tackled easily in the $d\to\infty$ limit is
the Gutzwiller state~\cite{Gutzwiller1963,Gutzwiller1965}
\begin{equation}
\left|G\right>=\exp\Bigg(-\alpha\sum_j n_{j\uparrow} n_{j\downarrow}\Bigg)\left|F\right>,\label{eq:Gutzwiller}
\end{equation}
that employs the local charge operators $n_{i\sigma}$ on sites $i$
to suppress local double occupancy in the reference Fermi sea
$\left|F\right>$ . 
The Gutzwiller state is very efficient
at creating correlations, and provides a qualitative picture of the Mott
transition~\cite{Metzner1989b}.
However, the Gutzwiller state provides only a crude description of the Fermi liquid phase.
Its single-particle momentum distribution has a flat step profile 
$n_{\bm k}=\left[1-Z\,\text{sign}(\en_{\bm k})\right]/2$, with $Z$ the quasiparticle weight 
and $\en_{\bm k}$ the electronic dispersion relation~\cite{Vollhardt1984,Metzner1989b}.
This defect of the Gutzwiller state can be traced to the fact that the operator in Eq.~(\ref{eq:Gutzwiller})
has momentum-independent amplitudes when expressed in Fourier space (apart from total momentum
conservation).
The proper physical behavior is easily seen from the half-filled ground state of the Hubbard model, 
expanded to first order in the interaction $U$:
\begin{align}
&\left|\text{GS}\right>\simeq\left|F\right>\nonumber
- U\! \int \prod_{l=1}^4 \frac{d^dk_{l}}{(2\pi)^d}\;
\theta(\en_{\bm k_1})\,\theta(\en_{\bm k_2})\theta(-\en_{\bm k_3})\theta(-\en_{\bm k_4})\\
&\times
\delta(\bm k_1+\bm k_2-\bm k_3-\bm k_4)\nonumber
T(\en_{\bm k_1},\en_{\bm k_2},\en_{\bm k_3},\en_{\bm k_4})\\
&\times
c_{\bm k_1\uparrow}^\dagger c_{\bm k_2\downarrow}^\dagger c_{\bm k_3\downarrow} c_{\bm k_4\uparrow}\left|F\right>,\label{eq:pert}  
\end{align}
where the perturbative amplitude
\begin{equation}
T(\en_{\bm k_1},\en_{\bm k_2},\en_{\bm k_3},\en_{\bm k_4})
=\frac{1}{|\en_{\bm k_3}|+|\en_{\bm k_4}|+|\en_{\bm k_1}|+|\en_{\bm k_2}|},
\label{eq:pertamp}
\end{equation}
indeed suppresses the participation of orbitals far from the Fermi energy through the large energy denominator.
A well-known trick is to introduce a Laplace transformation
that facilitates factoring the energy denominator into a product of momentum integrals that can be done indepedently
of each other
$T(\en_{\bm k_1},\en_{\bm k_2},\en_{\bm k_3},\en_{\bm k_4}) \!\!=\!\! \int_0^{+\infty} d\tau 
e^{-\tau(|\en_{\bm k_3}|+|\en_{\bm k_4}|+|\en_{\bm k_1}|+|\en_{\bm k_2}|)}$, 
leading to an imaginary time formulation.
With the goal of keeping a compact wave function description, 
one may make a ``rank one'' approximation of the Laplace integral 
\begin{equation}
T(\en_{\bm k_1},\en_{\bm k_2},\en_{\bm k_3},\en_{\bm k_4}) \simeq
\prod_{l=1}^4F(\en_{\bm k_l}),
\label{eq:singleenv}
\end{equation}
employing a single envelope $F(\en_{\bm k})$ that decays appropriately away from the Fermi level.
We comment on the accuracy of this approximation and its systematic improvement in Sec.~\ref{sec:friendly} below.
Expressing $F({\bm k}) = \sum_j e^{-i \bm k \cdot \bm r_j}$ in Eq.~(\ref{eq:pert}),
the perturbative wave function now takes the form:
\begin{eqnarray}
\nonumber
\left|\text{GS}\right> &\simeq& \left|F\right>
 - U \sum_j
\prod_{l=1}^4 \int \frac{d^dk_{l}}{(2\pi)^d}\;
\theta(\rho_l \en_{\bm k_l}) F(\en_{\bm k_l}) \\
&& \times e^{- i \rho_l \bm k_l \cdot \bm r_j}
c_{\bm k_1\uparrow}^{\rho_l} \left|F\right>,
\label{eq:pertloc} 
\end{eqnarray}
where $\rho_l=+$ (presence of $\dagger$ for the creation operators) if $l=1,2$ and
$\rho_l=-$ (absence of $\dagger$ for the destruction operators) if $l=3,4$.
This expression is the starting point of a coupled cluster construction that is
amenable to a large $d$ limit.

\subsection{Envelope particle-hole fermions}

Inspired by expression~(\ref{eq:pertloc}) for the perturbative state under the
the single envelope approximation~(\ref{eq:singleenv}),
we start by defining a set of $N_F\geq1$ particle ($+$) and $N_F\geq1$ hole ($-$) envelope
fermionic creation operators with spin polarization $\sigma\in\left\{\uparrow,\downarrow\right\}$,
for every lattice site $\bm r_m$:    
\begin{eqnarray}
    \psi_{m\sigma\alpha+}^\dagger\nonumber &=&
    \int\prod_{a=1}^d \frac{dk_a}{\sqrt{2\pi}} \,\theta(\en_{\bm k})   
    F_\alpha(\en_{\bm k})e^{-i\bm k\cdot\bm r_m}c_{\bm k\sigma}^\dagger,\nonumber\\
    \psi_{m\sigma\alpha-}^\dagger&=&
    \int\prod_{a=1}^d \frac{dk_a}{\sqrt{2\pi}}\,\theta(-\en_{\bm k})
    F_\alpha(\en_{\bm k})e^{i\bm k\cdot\bm r_m}c_{\bm k\sigma}.
    \label{eq:psi}
\end{eqnarray}
Here we refer to $\alpha\in\{1\ldots N_F\}$ as the flavor index. Throughout, the indices of these
operators are written in the order: site, spin, flavor, particle-hole ($\rho=\pm$), with the particle-hole index last.
Propagators and kernels, which are independent of spin in a paramagnetic state, carry the order: site, flavor, 
and particle-hole.
As we work in the particle-hole representation, all annihilation operators 
$\psi_{m\sigma\alpha\rho}$ now destroy the Fermi sea $\left|F\right>$. 
Different flavors are distinguished by different envelope functions $F_\alpha(\en)$ that 
determine the participation of orbitals with kinetic energy $\en$ in the flavor $\alpha$. 
Since the local single-particle basis contains only one orbital per spin per lattice site, 
the set $\left\{\psi^\dagger_{m\sigma\alpha+},\psi_{m\sigma\alpha-}|\alpha=1,\ldots,N_F;m=1,2,\ldots\right\}$ 
with $m$ running over all lattice sites, 
is associated with an overcomplete single-particle basis.

While $\psi_{m\sigma\alpha\rho}$ is not completely localized to site $m$, we will now prove that its amplitudes on sites
$n\ne m$ decay exponentially fast with distance between the two sites. 
Indeed, expressing $\psi^\dagger_{m\sigma\alpha+}$ in terms of 
the usual position basis operators
\begin{equation}
c_{i\sigma}^\dagger=\int \prod_{a=1}^d \frac{dk_a}{\sqrt{2\pi}} \,e^{-i\bm k\cdot \bm r_i}c_{\bm k \sigma}^\dagger,\label{eq:local}
\end{equation} 
yields
\begin{equation}
\psi_{m\sigma\alpha+}^\dagger=\sum_{i}\int d\en\,\theta(\en)F_\alpha(\en)g_\en(\bm r_m - \bm r_i)c_{i\sigma}^\dagger,\label{eq:psiitolocal}
\end{equation}
and similarly for $\psi_{m\sigma\alpha-}^\dagger$, with
the spatial correlator:
\begin{eqnarray}
g_\en(\bm r)
&=&\int\prod_{a=1}^d \frac{dk_a}{2\pi}\,\delta(\en-\en_{\bm k})e^{i\bm k\cdot\bm r}\nonumber\\
&=&\int \frac{d\tau}{2\pi}e^{i\tau\en}
\prod_{a=1}^d \int \frac{dk_a}{2\pi}\,e^{i \frac{2 t^*\tau}{\sqrt{d}}\cos(k)+i k_a r_{a}}\nonumber\\
&=&\int \frac{d\tau}{2\pi}e^{i\tau\en}
\prod_{a=1}^d i^{\left|r_{a}\right|}J_{\left|r_{a}\right|}\left(\frac{2 t^*\tau}{\sqrt{d}}\right).\label{eq:defsmallg}
\nonumber
\end{eqnarray}
The asymptotic behavior $J_n(2x)\sim x^n/n!$ of Bessel functions then
causes the spatial correlators $g_\en(\bm r_m - \bm r_i)$ to decay at least as fast as 
$d^{-|\bm r_m-\bm r_i|/2}$, in terms of 
the ``Manhattan" distance between sites.

\subsection{$d\to\infty$ friendly coupled cluster states}
\label{sec:friendly}
Equiped with the envelope fermions~(\ref{eq:psi}), and with the $d\to\infty$ limit in mind, 
we construct a family of coupled cluster states~\cite{Coester1958,Cizek1966,BartlettMusial2007},
based on treating the operators (\ref{eq:psi}) associated with all sites ${\bm r}_j$ on equal footing:
\begin{equation}
\left|\Psi\right>=\exp\Bigg(-\sum_j T_j^\dagger\Bigg)\left|F\right>,\label{eq:state}
\end{equation}
with the cluster operator for doubles excitations:
\begin{align}
&T_j^\dagger=\sum_{\alpha\beta\gamma\delta} A_{\alpha\beta\gamma\delta}\,\psi_{j\uparrow\alpha+}^\dagger\psi_{j\downarrow\beta+}^\dagger
\psi_{j\downarrow\gamma-}^\dagger\psi_{j\uparrow\delta-}^\dagger\nonumber\\
&
+\!\!\!\sum_{\alpha<\beta,\gamma>\delta} 
B_{\alpha\beta\delta\gamma}\sum_{\sigma\in\{\uparrow,\downarrow\}}
\psi_{j\sigma\alpha+}^\dagger\psi_{j\sigma\beta+}^\dagger\psi_{j\sigma\gamma-}^\dagger\psi_{j\sigma\delta-}^\dagger.\label{eq:clusterop}
\end{align}
A crucial point about coupled cluster states is that
all the terms in the summed argument of the exponential, being products of even numbers of creation operators, commute, and hence the exponential
factors into a product of exponentials. 
This particular ansatz is also motivated by our recent understanding (for 1D lattices) that uniform lattice
translations of local cluster operators offer drastic simplifications more general coupled
cluster wavefunctions~\cite{1dCoupledCluster}. The terms in $T_j^\dagger$ proportional to  
$A_{\alpha\beta\gamma\delta}$ excite one spin-up and one spin-down particle out of the Fermi sea,
while terms proportional to $B_{\alpha\beta\delta\gamma}$ excite two particles with the same spin polarization.
Due to Pauli principle, the `$B$' process is absent if we stick to a single flavor.
Note also that taking $A_{\alpha\beta\gamma\delta}$ and $B_{\alpha\beta\delta\gamma}$ independent of lattice site index $m$
ensures that $\left|\Psi\right>$ is lattice-translation invariant, and thus a zero-eigenstate of the total Bloch momentum.

The parametrization of the local cluster operator~(\ref{eq:clusterop}) is more general than it superficially
appears: By inserting the form of the envelope fermions~(\ref{eq:psi}), we can re-express 
the $A$ amplitude term of the cluster operator in momentum space (with
momentum conservation and occupation number restrictions implicit) as:
\begin{equation}
T_{\bm k_1 \bm k_2 \bm k_3 \bm k_4} = \sum_{\alpha\beta\gamma\delta} A_{\alpha\beta\gamma\delta} 
F_\alpha(\en_{\bm k_1})
F_\alpha(\en_{\bm k_2})
F_\alpha(\en_{\bm k_3})
F_\alpha(\en_{\bm k_4}),
\end{equation}
and similarly for the $B$ amplitude in~(\ref{eq:clusterop}). 
This corresponds precisely to the Tucker decomposition~\cite{TuckerCoupledCluster} 
of any tensor $T_{\bm k_1 \bm k_2 \bm k_3 \bm k_4}$ that, apart from momentum conservation, depends on momenta through the dispersion relation $\en_{\bm k}$ only, 
so that truncating to a small number $N_F$
of flavors amounts to a systematic low-rank approximation. 
For the first order energy denominator (\ref{eq:pertamp}) one can check that convergence is exponential in $N_F$. 
Since by construction our ansatz includes  lowest order perturbation theory of the Hubbard model,
we expect to obtain a realistic Fermi liquid ground state at weak coupling, with a better
description of the momentum distribution function than in the Gutzwiller state, thanks to the envelope
functions $F_\alpha(\varepsilon)$.
It is important to point out that unlike the first order state correction in Rayleigh-Schr\"odinger perturbation
theory~(\ref{eq:pert}), that has only two particle hole pairs, our trial state is size-consistent, i.e. 
it possesses a finite density of particle hole pairs. What is not yet clear is how much correlations
can be held in these coupled cluster states, especially in the regime near the Mott transition.

\section{Coupled cluster evaluation in the limit of infinite dimension}
\label{sec:eval}
\subsection{Static observables as Grassmann integrals}
Our framework to deal with the coupled cluster state~(\ref{eq:state})-(\ref{eq:clusterop}) draws on the mature understanding of 
Hubbard model diagrammatics in large spatial dimensions~\cite{Metzner1989b,Muller-Hartmann1989,Muller-Hartmann1989b,Georges1996}, that
forms the foundation of DMFT. The static expectation value in the free Fermi sea
\begin{equation}
[G_0]_{m\alpha\rho,n\beta\rho}=\left<F\right|\psi_{m\sigma\alpha\rho}\psi_{n\sigma\beta\rho}^\dagger\left|F\right>
\label{eq:zeroordergf}
\end{equation}
plays in our wave function theory the analogous role of the bare single-particle propagator in DMFT.
In the paramagnetic phase, spin indices can be dropped in observables, and in general the particle-hole
index $\rho=\pm$ is conserved.
We therefore refer to this static expectation value as the zero-order Green function. We stress however at
the outset that our formalism is not dynamic. We confine ourselves to the calculation of trial
ground states and expectation values, notwithstanding our borrowing of Green function terminology
like ``propagator", ``self-energy", or ``action".
From Eq.~(\ref{eq:psi}), we obtain:
\begin{equation}
[G_0]_{m\alpha\rho,n\beta\rho}=\int d\en\,\theta(\rho\en)F_\alpha(\en)^*F_\beta(\en)g_\en(\bm r_m - \bm r_n).\label{eq:Gitog}
\end{equation} 
By virtue of Eq.~(\ref{eq:defsmallg}), $[G_0]_{m\alpha\rho,n\beta\rho}$ decays at least as fast as $d^{-|\bm r_m-\bm r_n|/2}$, 
similar to the behavior of the dynamical single-particle Green functions at the heart of 
Hubbard model diagrammatics in large spatial dimensions.

In fact, the interpretation of $[G_0]_{m\alpha\rho,n\beta\rho}$ as a propagator becomes very natural
after the following observation. We first note that, by virtue of Wick's theorem, expectation values of
anti-normal ordered products of our fermion operators $\psi_{m\sigma\alpha\rho}$ and
$\psi_{n\sigma'\beta\rho'}^\dagger$ with respect to fermi sea $\left|F\right>$
can be expressed as determinants of matrices whose entries are
$[G_0]_{m\alpha\rho,n\beta\rho}$. Swapping the site, spin, flavor and particle-hole indices for a
single vectorial index that we denote $\bm m$, we get the general expectation values:
\begin{eqnarray}
&&\left<F\right|\psi_{\bm m_r}\ldots \psi_{\bm m_r} \psi_{\bm n_1}^\dagger\ldots \psi_{\bm n_s}^\dagger\left|F\right>\nonumber\\
&=&\delta_{rs}\,\text{Det}Q(\bm m_1 \ldots\bm m_r|\bm n_1\ldots \bm n_r),\label{eq:Wick}
\end{eqnarray}
where
\begin{equation}
[Q(\bm m_1 \ldots\bm m_r|\bm n_1\ldots \bm n_r)]_{ij}=[G_0]_{\bm m_i \bm n_j}.\label{eq:defq}
\end{equation}
Next, we recall that Gaussian integrals of Grassmann variables also produce determinants~\cite{Negele1988}. We then
introduce independent Grassmann numbers $\phi_{m\sigma\alpha\rho}\leftrightarrow \psi_{m\sigma\alpha\rho}$ and
$\bar \phi_{j\sigma'\alpha'\rho'}\leftrightarrow\psi_{n\sigma'\beta\rho'}^\dagger$ associated with our fermion operators.
By virtue of Eq.~(\ref{eq:Wick}) we can represent the expectation value as a Grassmann integral
\begin{eqnarray}
&&\left<F\right|\psi_{\bm m_r}\ldots \psi_{\bm m_r} \psi_{\bm n_1}^\dagger\ldots \psi_{\bm n_s}^\dagger\left|F\right>\nonumber\\
&=&\frac{\int\mathcal{D}[\bar\phi,\phi]\,\phi_{\bm m_1}\ldots \phi_{\bm m_r} \bar \phi_{\bm n_1}\ldots \bar \phi_{\bm n_s}e^{-\bar\phi G_0^{-1}\phi}}{\int\mathcal{D}[\bar\phi,\phi]\,e^{-\bar\phi G_0^{-1}\phi}},
\label{eq:Grassmann}
\end{eqnarray}
with the measure
\begin{equation}
\mathcal{D}[\bar\phi,\phi]=\prod_{\rho\sigma\alpha m}d\bar\phi_{m\sigma\alpha\rho}d\phi_{m\sigma\alpha\rho},\label{eq:measure}   
\end{equation}
and the free Gaussian kernel
\begin{equation}
\bar\phi G_0^{-1}\phi=\sum_{\rho\sigma}\sum_{m\alpha n\beta}\bar\phi_{m\sigma\alpha\rho}[G_0^{-1}]_{m\alpha\rho,n\beta\rho}\phi_{n\sigma\beta\rho}.\label{eq:quadaction}   
\end{equation}
A technical point should be mentioned here. The matrix $G_0$ is in fact not invertible. One may however
introduce a regularized matrix, e.g. $G_0^{(r)}=G_0+i I\delta$, which is invertible, and send the regularization 
parameter $\delta$ to zero at the end of the calculation. One finds regular expressions for observable quantities.
 This can essentially be seen by inspection of final results,
which are all expressed in terms of matrix elements of $G_0$, and are regular without any need to cancel singularities in denominators 
against zeros in numerators. Fundamentally we believe 
the reason for this is that it is rather the cofactor matrix $\text{Det}(G_0)\,G_0^{-1}$ than
$G_0^{-1}$ itself that determines physical quantities, although our calculations only confirm this rigorously
in the limit $d\to\infty$, since we drop all $1/d$ and higher order corrections before sending $\delta\to0$.
Since we never explicitly need to invoke the regularized form of $G_0$ or take the $\delta\to0$ limit
in any nontrivial way, we keep writing $G_0^{-1}$ in what follows, implicitly
assuming a regularization in intermediate results. 

Let $O[\psi,\psi^\dagger]$ be any expression in which the operators $\psi_{m\sigma\alpha\rho}$ and
$\psi_{n\sigma'\beta\rho'}^\dagger$ appear in anti-normal ordered form. Using the Grassmann representation
that we established above, we can write for its expectation value with respect to the coupled cluster state
$\left|\Psi\right>$ of Eq.~(\ref{eq:state}):  
\begin{align}
&\frac{\left<\Psi\right|O[\psi,\psi^\dagger]\left|\Psi\right>}{\left<\Psi|\Psi\right>}\nonumber\\
&=\frac{\int\mathcal{D}[\bar\phi,\phi]\,O[\phi,\bar\phi]e^{-\sum_j(T_j+\bar T_j)-\bar\phi G_0^{-1}\phi}}{\int\mathcal{D}[\bar\phi,\phi]\,e^{-\sum_j(T_j+\bar T_j)-\bar\phi G_0^{-1}\phi}}.\label{eq:expvalintegral}
\end{align}
Here $\bar T_j + T_j$ represents the operator $T_j^\dagger + T_j$ from Eq.~(\ref{eq:clusterop}), with 
the replacement $\psi_{j\sigma\alpha\rho},\,\psi_{j\sigma\alpha\rho}^\dagger\rightarrow
\phi_{j\sigma\alpha\rho},\,\bar \phi_{j\sigma\alpha\rho}$. 
Note that $\bar T_j$ and $T_j$ are Grassmann-even and therefore commute.

\subsection{Cavity construction for wave functions}
In the $d\to\infty$ limit, the Grassmann integral formulation allows us to construct an embedding theory
similar to DMFT, for expectation values of local operators defined on a single site $o$,
corresponding to say the origin $\bm r_o = \bm 0$.
In other words, the effect of correlations in the rest of the system on the expectation value $O[\psi_o,\psi_o^\dagger]$
of the local operator at the origin is exactly accounted for in the $d\to\infty$ limit, by adding a quadratic term to the local action.
The construction works as follows. In Eq.~(\ref{eq:expvalintegral}), we group the integral over Grassmann numbers associated
with sites other than $o$:
\begin{align}
&\frac{\left<\Psi\right|O[\psi_o,\psi_o^\dagger]\left|\Psi\right>}{\left<\Psi|\Psi\right>}\nonumber\\
&=\frac{\int\mathcal{D}[\bar\phi_o,\phi_o]\,O[\phi_o,\bar\phi_o]e^{-T_o-\bar T_o-\bar\phi_o [G_0^{-1}]_{o o}\phi_o}\mathcal I[\phi_o,\bar \phi_o]}{\int\mathcal{D}[\bar\phi_o,\phi_o]\,e^{-T_o-\bar T_o-\bar\phi_o [G_0^{-1}]_{o o}\phi_o}\mathcal I[\phi_o,\bar \phi_o]},\label{eq:localint1}
\end{align}
into an ``influence functional''
\begin{widetext}
\begin{align}
\mathcal I[\phi_o,\bar \phi_o]=\frac{\int\prod_{j\ne o}\mathcal{D}[\bar\phi_j,\phi_j]\,e^{-\sum_{j\ne o}(T_j+\bar T_j+\bar\eta_j \phi_j+\bar\phi_j\eta_j)-\sum_{mn\ne o}\bar\phi_m [G_0^{-1}]_{m n}\phi_n}}{\int\prod_{j\ne o}\mathcal{D}[\bar\phi_j,\phi_j]\,e^{-\sum_{mn\ne o}\bar\phi_m [G_0^{-1}]_{m n}\phi_n}},\label{eq:influence}
\end{align}
\end{widetext}
where $\eta_j=[G_0^{-1}]_{j o}\phi_o$ and $\bar\eta_j=\bar\phi_o[G_0^{-1}]_{o j}$.
It is important to note here that the (regularized) bare propagator $G_0^{-1}$ has the same scaling with separation as $G_0$,
i.e. $[G_0^{-1}]_{m\alpha\rho,n\beta\rho}\sim d^{-|\bm r_m-\bm r_n|/2}$ -- one simply replaces
$\theta(\pm \en_{\bm k})F_\alpha(\en_{\bm k})^*F_\beta(\en_{\bm k})$ with
$\left[\theta(\pm \en_{\bm k})F_\alpha(\en_{\bm k})^*F_\beta(\en_{\bm k})+iF\right]^{-1}$ 
in Eq.~(\ref{eq:Gitog}).
Thus the Grassmann fields $\eta_j$ and $\bar \eta_j$ scale like $d^{-|\bm r_j|/2}$.

The quantities we are studying can be represented in a Feynman diagram expansion of
the static influence functional~(\ref{eq:influence}) in powers of the cluster vertices $T_j$ and
$\bar T_j$.
It is natural to use the position representation.
Directed lines representing $[G_0]_{m\alpha\rho,n\beta\rho}$ originate or terminate
at vertices. Internal vertices represent sources $\bar T_n$ where four lines originate, one for each
particle-hole and spin polarization combination, and sinks $T_m$ where similarly, four lines end.
External vertices are labeled by the site(s) and other indices of the operator whose expectation value
is being computed.

By virtue of the linked cluster theorem~\cite{Negele1988}, the log of the influence functional (\ref{eq:influence}) 
can be expressed formally in terms of a sum over all connected Green functions:
\begin{widetext}
\begin{equation}
    \ln\left(\mathcal I[\phi_o,\bar \phi_o]\right)=\sum_{\substack{p,q\ge 1}}\ \sum_{n_1\ldots n_p\ne o}\ \sum_{m_1\ldots m_q\ne o}\bar \eta_{n_p}\ldots \bar \eta_{n_1} \left[G^{(c)}_{[o]}\right]_{n_1,\ldots,n_p;\,m_q,\ldots,m_1}\eta_{m_1}\ldots\eta_{m_q}.\label{eq:linkedcluster}
\end{equation}
\end{widetext}
Above, $G^{(c)}_{[o]}$ is the multi-particle connected cavity Green function with $p$ external sinks
(at sites $n_1,\ldots,n_p$) and $q$ external sources (at sites $m_1,\ldots,m_q$); internal vertices are summed over all sites
except $o$, and indices other than site indices are suppressed. Since every propagator connects one source to one sink, while
each internal sink $T_m$ absorbs four lines and each internal source $\bar T_n$ emits four, $p-q$
must be a multiple of four.
In the derivation of the usual DMFT equations, a similar linked cluster expression is obtained involving
connected dynamical Green functions, with the same scaling of sources as $\eta_n\sim d^{-|\bm r_n|/2}$ 
and bare propagators as $d^{-|\bm r_m-\bm r_{n}|/2}$. By analogy, we can conclude without calculation
that only the $p=q=1$ term in Eq.~(\ref{eq:linkedcluster}) survives in the limit $d\to\infty$.
(See for instance the discussion below Eq.~(34) in the review article \cite{Georges1996}.)
Thus, in the limit $d\to\infty$, 
$\mathcal I[\phi_o,\bar \phi_o]$ becomes a Gaussian in $\bar \phi_o$ and $\phi_o$, which we write as 
\begin{equation}
\mathcal I[\phi_o,\bar \phi_o]=\exp\left[\bar\phi_o([G_0^{-1}]_{o o}-\mathcal{G}_0^{-1})\phi_o\right],\label{eq:gaussian}
\end{equation}
so that, by combining with Eq.~(\ref{eq:localint1}), the quadratic part of the effective action of the 
local fermion field has kernel $\mathcal{G}_0^{-1}$, which we call the cavity field by analogy to DMFT.
It is still defined in terms of the unknown one-particle cavity propagator
$[G^{(c)}_{[o]}]_{n,m}$ inside the influence functional $\mathcal I[\phi_o,\bar\phi_o]$ at this stage.)
We arrive at an exact effective description of operators associated with site $o$:
\begin{align}
&\frac{\left<\Psi\right|O[\psi_o,\psi_o^\dagger]\left|\Psi\right>}{\left<\Psi|\Psi\right>}\nonumber\\
&~~~~~~~~~~=\frac{\int\mathcal{D}[\bar\phi_o,\phi_o]\,O[\phi_o,\bar\phi_o]e^{-T_o-\bar T_o-\bar\phi_o \mathcal{G}_0^{-1} \phi_o}}{\int\mathcal{D}[\bar\phi_o,\phi_o]\,e^{-T_o-\bar T_o-\bar\phi_o \mathcal{G}_0^{-1}\phi_o}}.\label{eq:effectivetheory}
\end{align}
reminiscent of a quantum impurity embedded in a non-interacting environment, in that non-Gaussian
term in the integrals $T_o$ and $\bar T_o$ involve only Grassmann numbers on the local site $o$.
What makes the problem simple here, compared to DMFT, is that we are dealing with a finite number of Grassmann numbers
instead of fields depending on time (or frequency), and one can therefore perform the
integrals~(\ref{eq:effectivetheory}) by brute force, provided we
restrict ourselves to a modest number $N_F$ of flavors.

Performing the integrals in Eq.~(\ref{eq:effectivetheory}) however does not complete the calculation.
The matrix $\mathcal{G}_0^{-1}$ also needs to be determined. Rather than calculating the one-particle static Green's
function $G^{(c)}_{[o]}$ diagrammatically, one may follow the same self-consistent route as is done in DMFT.
Let us introduce the expectation value in the full coupled cluster state
\begin{equation}
G_{m\alpha\rho,n\beta\rho}
=\frac{\left<\Psi\right|\psi_{m\sigma\alpha\rho}\psi_{n\sigma\beta\rho}^\dagger\left|\Psi\right>}{\left<\Psi|\Psi\right>},    
\end{equation}
which plays the role of the interacting lattice Green function in DMFT.
The fact that our diagrammatic expansion of $G_{m\alpha\rho,n\beta\rho}$ has internal
``interaction'' vertices where four lines meet, together with the scaling of the bare ``propagator'' 
$[G_0]_{m\alpha\rho,n\beta\rho}\sim d^{-|\bm r_m-\bm r_n|/2}$, then results in amputated single-particle irreducible (1PI)
diagrams associated with $G$, that we denote $\Sigma$ by analogy to the DMFT self-energy, and which
are diagonal in site indices, i.e. written as a matrix equation:
\begin{equation}
G=(1-G_0\Sigma)^{-1}G_0=(G_0^{-1}-\Sigma),\label{eq:dyson}
\end{equation}
where
\begin{equation}
\Sigma_{m\alpha\rho,n\beta\rho}=\delta_{mn}\Sigma_{\alpha\rho,\beta\rho}.\label{eq:Sigma1}
\end{equation}
The diagonality of the ``self-energy'' in position representation follows from the fact that in any
two-point 1PI correlator of a $\phi^4$-like theory, the two external legs $m$ and $n$ are connected by at least
three independent paths (i.e. paths that share no zero-order propagator lines). This implies
$\Sigma_{m n}\sim d^{-3|\bm r_m-\bm r_{n}|/2}$, while
summation over sites at fixed Manhattan distance $|\bm r_m-\bm r_n|$ only involves 
$~d^{|\bm r_m-\bm r_n|}$ terms, hence only local terms survive at $d\to\infty$. 
This was pointed out in for instance
Refs.\,\cite{Metzner1989,Metzner1989b,Muller-Hartmann1989}.  At this stage
$\Sigma_{\alpha\rho,\beta\rho}$ is not known, and the brute force way of
calculating it requires summing all diagrams that cannot be separated into two pieces by
cutting one non-interacting propagator line.

The scaling $[G_0^{-1}]_{m\alpha\rho,n\beta\rho}\sim d^{-|\bm r_m-\bm r_n|/2}$ also implies 
that the cavity Green function $G_{[o]}^{(c)}$ is related 
to the interacting Green function $G$ in the same ways as in DMFT.
We explain this further in Appendix \ref{app:A} and show that the result
is an exact ``impurity''-like Dyson equation:
\begin{equation}
G_{o o}=(\mathcal{G}_0^{-1}-\Sigma)^{-1},\label{eq:selfcon1}
\end{equation}
between the interacting local Green function at site $o$, and the unknown cavity field $\mathcal{G}_0^{-1}$ and
``self-energy'' $\Sigma$, both viewed as matrices in particle-hole and flavor indices. 
i.e. from now on $\Sigma$ refers to the matrix on the right-hand side of Eq.~\ref{eq:Sigma1}. 
Eq.~(\ref{eq:effectivetheory}), 
with $O[\psi_o,\psi_o^\dagger]=\psi_{o\sigma\alpha\rho}\psi_{o\sigma\beta\rho}^\dagger$,
Eq.~(\ref{eq:dyson}), evaluated at site indices $o,o$, and Eq.~(\ref{eq:selfcon1}) then represent three equations, that
can be solved simultaneously to find the unknown quantities $G_{o o}$, $\mathcal{G}_0^{-1}$ and $\Sigma$, in
complete analogy to DMFT. Having determined $\mathcal{G}_0^{-1}$, one can then subsequently 
compute the expectation value of any anti-normal ordered operator involving fermion operators associated 
with site $o$. Having determined $\Sigma$, one can also compute the full interacting Green function 
via Eq.~\ref{eq:dyson}. As a result, one can compute expectation values of all operators that 
are quadratic in fermionic operators. 

\section{Single flavor doubles: closed-form results}
\label{sec:givendelta}
\subsection{Solution of the $d\to\infty$ equations}
\label{sec:soldoubles}
\begin{figure}[ht]
    \begin{center}
    \includegraphics[width=.99\columnwidth]{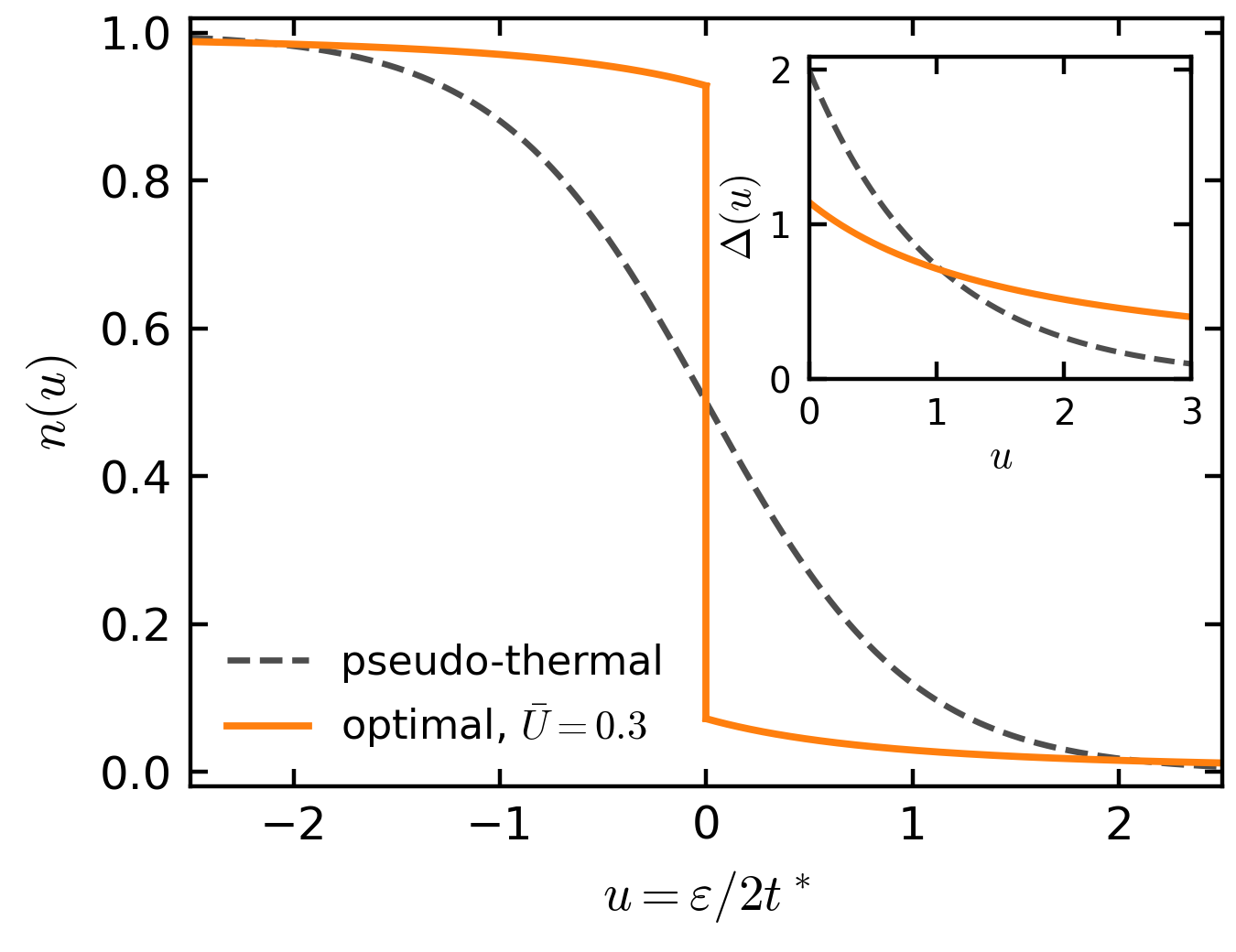}
    \end{center}
    \caption{{\bf Main panel:} The pseudo-thermal distribution $n(u)=e^{-2u}/(1+e^{-2u})$ produced by
    the scaled envelope $f(u)=e^{-u}$ is contrasted with the distribution found for the envelope
    that minimises the variational energy for the Hubbard model at interaction $U=0.3\times 16 t^*
    /\sqrt{\pi}=2.7 t^*$ where the coupled cluster state gives approximately the same variational 
    energy as the Gutzwiller state, see Sec.~\ref{sec:Hubbard} below.  
    {\bf Inset:} Comparison of the envelopes $F(\en)$ producing the two single-particle
    momentum distributions in the main panel.
    \label{fig:n}}
\end{figure}

With the general formalism established, we now consider the simplest example, namely that of a
coupled cluster doubles with single fermionic flavor, defined through a single
particle-hole symmetric envelope $F(\en)$. We will therefore drop the index $\alpha$
in all subsequent propagators.
Without loss of generality we may set the single coefficient $A$ in $T^\dagger$ equal to $1$.
(Positive sign of $A$ ensures suppression of double occupancy).
The non-interacting Green function, and the interacting Green function, expressed in terms of the 
``self-energy'' (\ref{eq:dyson}) then read respectively:
\begin{align}
&[G_0]_{m\rho,n\rho}=\int\prod_{a=1}^d \frac{dk_a}{2\pi}\,
    \theta(\rho\en_{\bm k})F(\en_{\bm k})^2 e^{i\rho\bm k\cdot(\bm r_m-\bm r_n)},\nonumber\\
&G_{m\rho,n\rho}=\int\prod_{a=1}^d \frac{dk_a}{2\pi}\,
    \theta(\rho\en_{\bm k})\frac{F(\en_{\bm k})^2}{1-F(\en_{\bm k})^2\Sigma} 
    e^{i\rho\bm k\cdot(\bm r_m-\bm r_n)}\nonumber,\\
    \label{eq:gf1}
\end{align}
where now $\Sigma$ is just a scalar due to the single flavor considered, and to particle-hole
symmetry. We emphasize again that $\Sigma$ is not a self-energy in the usual many-body sense, but
rather the irreducible correction to the one-particle correlator $G$.
The free density of states (per unit volume per spin) associated with the energy dispersion relation
(\ref{eq:dispersion}) in the $d\to\infty$ limit is
$P(\en/2t^*)/2t^*$~\cite{Metzner1989b,Muller-Hartmann1989}
where
\begin{equation}
P(u)=\frac{1}{\sqrt{\pi}}e^{-u^2}.
\end{equation}
From now on, we will work with the dimensionless energy $u=\en/2t^*$. We view $F$ as a function of $u$
and from Eq.~(\ref{eq:gf1}) obtain expressions
\begin{align}
[G_0]_{o o}&=\int_0^\infty du\,P(u)F(u)^2,\nonumber\\
G_{o o}&=\int_0^\infty du\,P(u)\frac{F(u)^2}{1-F(u)^2\Sigma},\label{eq:localgf1}
\end{align}
for the local (i.e. at site $o$) non-interacting and interacting Green functions.
We perform the Grassmann integral over the eight Grassmann numbers $\phi_{o\sigma\rho}$ and
$\bar \phi_{o\sigma\rho}$ in Eq.~(\ref{eq:effectivetheory}) to obtain an expression for the
interacting Green function $G_{o o}$ in terms of the Weiss field $\mathcal{G}_0$, which in the case of 
a single species, is a scalar (i.e. a $1\times 1$ matrix), given by:
\begin{eqnarray}
G_{o o}&=&\frac{\int\mathcal{D}[\bar\phi_o,\phi_o]\,\phi_{o\sigma\rho}\bar\phi_{o\sigma\rho}e^{-\bar\phi_o \mathcal{G}_0^{-1} \phi_o}}{\int\mathcal{D}[\bar\phi_o,\phi_o]\,(1+\prod_{\rho,\sigma}\phi_{o\sigma\rho}\bar\phi_{o\sigma\rho})e^{-\bar\phi_o \mathcal{G}_0^{-1}\phi_o}}\nonumber\\
&=&\frac{\mathcal{G}_0}{1+\mathcal{G}_0^{4}}.\label{eq:gf1itoK}
\end{eqnarray}
To arrive at this expression from Eq.~(\ref{eq:effectivetheory}), we used the fact that
the Grassmannian terms $T_o$ and $\bar T_o$ are nilpotent, so that $e^{-T_o-\bar T_o}=1-T_o-\bar T_o+T_o\bar T_o$, 
with $T_o\bar T_o=\prod_{\rho\sigma}\phi_{o\sigma\rho}\bar\phi_{o\sigma\rho}$. 
The terms linear in $T_o$ or $\bar T_o$ do not contribute, since the Gaussian integral is non-zero
only when pairing each $\phi_{o\sigma\rho}$ with $\bar\phi_{o\sigma\rho}$. 
In the numerator, the term $T_o\bar T_o$ also vanishes, since it would contain $\phi_{o\sigma\rho}$ 
twice due to the fields generating the propagator.
Equating Eq.~(\ref{eq:gf1itoK}) to Eq.~(\ref{eq:selfcon1}) allows us to find the ``self-energy''
$\Sigma$ in terms of $\mathcal{G}_0^{-1}$:
\begin{equation}
\frac{\mathcal{G}_0}{1+\mathcal{G}_0^{4}}=\frac{1}{\mathcal{G}_0^{-1}-\Sigma}\implies\Sigma=-\mathcal{G}_0^{3},\label{eq:SigmaitoK}
\end{equation}
formally analogous to the sunrise diagram of the dynamical impurity self-energy in DMFT.
Finally, equating Eq.~(\ref{eq:gf1itoK}) to Eq.~(\ref{eq:localgf1}) is the analog of the DMFT
self-consistency, and using expression~(\ref{eq:SigmaitoK}) for the ``self-energy'', we find a
self-consistent condition for the Weiss field $\mathcal{G}_0$:
\begin{equation}
\frac{\mathcal{G}_0}{1+\mathcal{G}_0^{4}}=\int_0^\infty du\,P(u)\frac{F(u)^2}{1+F(u)^2 \mathcal{G}_0^{3}}.\label{eq:self1}
\end{equation}
We may as well multiply Eq.~(\ref{eq:self1}) by $\mathcal{G}_0^{3}$, and define
\begin{equation}
f(u)=\mathcal{G}_0^{3/2}F(u).\label{eq:deff}
\end{equation} 
The normalized envelope function $f:[0,\infty)\to \mathbb{R}$ is the object that specifies the actual coupled
cluster state, and any function is a valid choice (it may also be optimized variationally on a given
Hamiltonian, as we will consider in Sec.~\ref{sec:Hubbard}).
For a given profile of $f(u)$, Eq.~(\ref{eq:gf1}) can then be Fourier transformed to obtain the momentum
distribution function
$n_{\bm k}=\left<\Psi\right|c_{\bm k\sigma}^\dagger c_{\bm k\sigma}\left|\Psi\right>/\left<\Psi\right|\left.\Psi\right>$. 
One finds
\begin{equation}
    n_{\bm k}=
    \left\{
    \begin{array}{ll}
        \frac{1}{1+f\left(-\frac{\en_{\bm k}}{2t^*}\right)^2} & \text{if } \en_{\bm k}<0\\
        \frac{f\left(\frac{\en_{\bm k}}{2t^*}\right)^2}{1+f\left(\frac{\en_{\bm k}}{2t^*}\right)^2} & \text{if } \en_{\bm k}>0
    \end{array}
    \right.\label{eq:nk}
\end{equation}
Remarkably, this provides a way to construct a correlated state with any desired momentum distribution of the form $n_{\bm k}=n(\en_{\bm k})$
consistent with particle-hole symmetry $n(-\en)=1-n(\en)$. Working again with dimensionless energy
$u$, one can re-express (\ref{eq:self1}) from (\ref{eq:deff}) and (\ref{eq:nk}), and we find:
\begin{equation}
\frac{1}{1+\mathcal{G}_0^{-4}}=\int_0^\infty du\, P(u) n(u),\label{eq:Kiton}
\end{equation}
i.e. $1/(1+\mathcal{G}_0^{-4})$ is the density per spin of particle-hole excitations, and an arbitrary distribution
$n(u)$ can be produced by the envelope
\begin{equation}
F(u)=\mathcal{G}_0^{-3/2}\sqrt{\frac{n(u)}{1-n(u)}}.\label{eq:Diton}
\end{equation} 
If $f(0)\ne 1$, then the momentum distribution has a jump at the Fermi surface. 
However, the choice 
\begin{equation}
f(u)^2=\frac{1-2 \tilde f(u)}{1+2 \tilde f(u)}
\end{equation}
where $\tilde f$ is a smooth real antisymmetric function bounded from above by $1/2$, leads to a
smooth momentum distribution $n(u)=1/2-\tilde f(u)$.  For instance, the ``pseudo-thermal" distribution
$n(u)=e^{-\beta u}/(1+e^{-\beta u})$ corresponds to $f(u)=e^{-\beta u/2}$. In Figure\,\ref{fig:n}, 
we illustrate this and compare the envelope $F(u)=\mathcal{G}_0^{-3/2}f(u)$ of this distribution to that found
by minimizing the variational energy of the Hubbard model, see Sec.~\ref{sec:Hubbard} below for
details of the latter calculation.

\subsection{Parent Hamiltonian}
It is generally possible to construct a parent Hamiltonian whose exact ground state is
$\left|\Psi\right>$. The argument runs along the same lines as the gossamer metal construction in
\cite{Kollar2003}, that yields a parent Hamiltonian for the Gutzwiller state, and which has been
extended to coupled cluster states recently~\cite{1dCoupledCluster}. We set
$T^\dagger=\sum_j T_j^\dagger$.  Taking as departure point that $c_{\bm k\sigma}^\dagger\left|F\right>=0$ for $\en_{\bm k}<0$ and 
$c_{\bm k\sigma}\left|F\right>=0$ for $\en_{\bm k}>0$, we define operators
\begin{align}
&h_{\bm k\sigma}=e^{-T^\dagger}c_{\bm k\sigma}^\dagger e^{T^\dagger}=c_{\bm k\sigma}^\dagger+\left[c_{\bm k\sigma}^\dagger,T^\dagger\right]\text{ for }\en_{\bm k<0},\nonumber\\
&p_{\bm k\sigma}=e^{-T^\dagger}c_{\bm k\sigma}e^{T^\dagger}=c_{\bm k\sigma}+\left[c_{\bm k\sigma},T^\dagger\right]\text{ for }\en_{\bm k>0},\label{eq:ph}
\end{align}
that trivially annihilate the state $\left|\Psi\right>$ by construction. 
Note that since $e^{T^\dagger}$ is not unitary,
$h_{\bm k\sigma}$ and $p_{\bm k\sigma}$ do not obey the usual anti-commutation relations with their Hermitian
conjugates. Nonetheless, we can define a positive semi-definite Hermitian parent Hamiltonian:
\begin{equation}
H_\text{p}=\sum_\sigma\int\!\prod_{a=1}^{d}dk_a\!\left[-\en_{\bm k}\theta(-\en_{\bm k})h_{\bm k\sigma}^\dagger h_{\bm k\sigma}+\en_{\bm k}\theta(\en_{\bm k})p_{\bm k\sigma}^\dagger p_{\bm k\sigma}\right]\label{eq:Hpdeff}
\end{equation}
for which $\left|\Psi\right>$ is the unique (zero energy) ground state by construction.
Writing $h_{\bm k\sigma}$ and $p_{\bm k\sigma}$ out explicitly and performing momentum integrals, we obtain
\begin{eqnarray}
H_\text{p}
&=&\sum_\sigma\int\prod_{a=1}^{d}dk_a\,\left[-\en_{\bm k}\theta(-\en_{\bm k})c_{\bm k\sigma} c_{\bm k\sigma}^\dagger
  +\en_{\bm k}\theta(\en_{\bm k})c_{\bm k\sigma}^\dagger c_{\bm k\sigma}\right]\nonumber\\
&&-\frac{t^*}{\sqrt{d}}\sum_{\rho\sigma m}\sum_{n\in\text{NN}_m}
  \psi_{n\sigma\rho}^\dagger\,\psi_{m\tilde\sigma\rho}^\dagger\,\psi_{m\tilde\sigma\tilde\rho}^\dagger\,\psi_{m\sigma\tilde\rho}^\dagger
  +\text{h.c.}\nonumber\\
&&+\sum_{\rho\sigma mn}V_{m-n,\tilde\rho}\;
  \psi_{m\tilde\sigma\rho}\psi_{n\tilde\sigma\rho}^\dagger\,\psi_{m\tilde\sigma\tilde\rho}\psi_{n\tilde\sigma\tilde\rho}^\dagger\,
  \psi_{m\sigma\rho}\psi_{n\sigma\rho}^\dagger ,
\label{eq:Hp2}
\end{eqnarray}
where
\begin{equation}
V_{j,\rho}=-\rho\,\frac{t^*}{\sqrt{d}}\sum_{l\in\text{NN}_j} [G_0]_{l\rho,0\rho},\label{eq:vdef}
\end{equation}
and we use the notation $\tilde \uparrow =\downarrow$, $\tilde \downarrow =\uparrow$
to indicate the opposite spin polarization to a given one, $\tilde + = -$ and $\tilde - =+$
for the particle/hole index as well, and $\text{NN}_m$ denotes the set of sites that are nearest 
neighbors of site $m$.
The parent Hamiltonian has the same one-body term as free fermions performing nearest neighbor hopping on the
hypercubic lattice. Additionally, it has correlated hopping terms involving two- and three-body processes.
The two- and three-body terms in $H_\text{p}$ inherit the following property from Eqs. (\ref{eq:psiitolocal}) and (\ref{eq:Gitog}).
When written in the fully local basis $c_{m\sigma}$, any term involving $c_{m\sigma}^{(\dagger)}$ and
$c_{n\sigma}^{(\dagger)}$, decays at least as fast as $d^{-|\bm r_m-\bm r_n|/2}$ between sites.
An important difference between the two-body term in (\ref{eq:Hp2}) and a Hubbard-like interactions is that
$\psi_{n\sigma\rho}^\dagger\,\psi_{m\tilde\sigma\rho}^\dagger\,\psi_{m\tilde\sigma\tilde\rho}^\dagger\,\psi_{m\sigma\tilde\rho}^\dagger$ 
only excites two particle-hole pairs, one spin-up, one spin-down, and
never de-excites particle-hole pairs, as $n_{m\uparrow}n_{\downarrow}$ does.

\subsection{A gapless state with zero quasiparticle weight}
Our gossamer metal construction also provides a proof that for $H_\text{p}$, the energy cost of
adding or removing a particle from the ground state $\left|\Psi\right>$ vanishes, i.e. that the
single-particle tunnelling spectrum is gapless. Consider the state 
$c_{\bm k\sigma}^\dagger|\Psi\rangle$ adding a particle, which is of course not an eigenstate of the parent
Hamiltonian $H_\text{p}$, but a variational upper bound on its excitation spectrum.
Let us take first $\bm k$ with $\en_{\bm k}>0$, so that $c_{\bm k\sigma}^\dagger$ does not annihilate 
the Fermi sea $\left|F\right>$. Since $[T^\dagger,c_{\bm k\sigma}^\dagger]=0$, we have:
\begin{equation} 
\{p_{\bm k\sigma},c_{\bm k'\sigma}^\dagger\}=\delta_{\bm k,\bm k'},\quad
\{h_{\bm k\sigma},c_{\bm k\sigma}^\dagger\}=0.  
\end{equation} 
Together with the fact that $h_{\bm p\sigma}$ and $p_{\bm p\sigma}$ annihilate $\left|\Psi\right>$, 
this implies that
\begin{equation}
E_\text{trial}(\bm k)=\frac{\left<\Psi\right|c_{\bm k\sigma} H_\text{p} c_{\bm k\sigma}^\dagger\left|\Psi\right>}{\left<\Psi\right|c_{\bm k\sigma} c_{\bm k\sigma}^\dagger\left|\Psi\right>}
=\frac{\en_{\bm k}}{1-n(\en_{\bm k})}.
\end{equation}
A similar result follows for the case of the removal of a particle, starting from a trial state 
$c_{\bm k\sigma}\left|\Psi\right>$ with $\en_{\bm k}<0$. Using the expression
(\ref{eq:nk}) for $n(\en_{\bm k})$ in terms of $f(|\en_{\bm k}|/2t^*)$, we get in both cases
\begin{equation}
E_\text{trial}(\bm k)=|\en_{\bm k}|\left[1+f\left(\frac{\en_{\bm k}}{2t^*}\right)^2\right].\label{eq:etrial}
\end{equation}
This upper variational bound on the true excitations vanishes at the Fermi energy, also when $n(\en)$ 
is smooth at the Fermi surface (namely when $f(0)=1$). 
Thus, there is no energy gap between $\left|\Psi\right>$ and the eigenstates of $H_\text{p}$ containing 
one extra or one less particle than $\left|\Psi\right>$.
We note that in the context of Gutzwiller trial states, the energy $E_\text{trial}(\bm k)$ has been 
interpreted as the quasi-particle dispersion relation in the Gutzwiller approximation for the Fermi-liquid 
hosted by a Hubbard model \cite{Bunemann2003}. Although we are not able to calculate dynamical response
coefficients such as the Drude weight, it seems likely that for $F(u)$ chosen to produce smooth 
$n(\en)$ (namely $Z=0$), the parent Hamiltonian $H_\text{p}$ realizes a gapless metal whose single-particle 
momentum distribution has no trace of a Fermi surface.

\section{Application to the half-filled Hubbard model}
\label{sec:Hubbard}

\subsection{Solution for the single flavor doubles}
Now we study the performance of the coupled cluster doubles state $\left|\Psi\right>$ as a
variational ground state approximation for the paramagnetic half-filled Hubbard model. In the
$U\lesssim t^*$ regime, this allows us to explore how controlling the momentum distribution improves
Fermi liquid-like properties compared to the Gutzwiller state. In the $U\gtrsim t^*$ regime, it
sheds light on the difficulties coupled cluster states have to capture strong correlations. In
particular, since evaluation of expectation values is exact, it identifies limits to the
expressivity of coupled cluster states, and separate these from failures of the usual approximations
made when evaluating coupled cluster states in quantum chemistry.

From the previous section~(\ref{sec:soldoubles}), we can obtain all one-particle expectation values
in the coupled cluster state, so that evaluation of the kinetic energy is straightforward.
In order to allow us to evaluate the double occupancy $\left<n_{o\uparrow}n_{o\downarrow}\right>$
on a given site $o$, we extend the framework of the previous section by introducing a
new auxilatory fermion species $\psi_{m\sigma 0\rho}$ associated with a flat envelope $F_0(\en)=1.$
The flavor index $\alpha=0,1$ now takes two values. We take $\alpha=1$ for the non-trivial envelope.
We work with the same $\left|\Psi\right>$ as before, that only involves the species
$\psi_{m\sigma 1\rho}$. The new species $\psi_{m\sigma 0\rho}$ only enters expressions for
expectation values by being inserted at external vertices. Nonetheless, to accommodate this in the
diagrammatic theory,
we must now endow Green functions with matrix structure, corresponding to the flavor index $\alpha\in\{0,1\}$.
We thus get the free propagator for the two fields:
\begin{equation}
[G_0]_{o o}=\int_0^\infty du\,P(u)
\left(
\begin{array}{cc}
    1&F(u)\\
    F(u)&F(u)^2
\end{array}
\right).\label{eq:G02x2}
\end{equation}
Since $\psi_{m\sigma 0\rho}$ is not part of the coupled cluster wave function, it never 
appear at internal vertices, so that only the $(1,1)$ component of the ``self-energy'' is nonzero:
\begin{equation}
\Sigma_{\alpha\beta}=\delta_{\alpha,1}\delta_{\beta,1}\Sigma.\label{eq:Sigma2x2}    
\end{equation}
Substituting this into the Dyson-like equation~(\ref{eq:dyson}) that relates the full $G$ 
to the bare $G_0$ and $\Sigma$, we get:
\begin{eqnarray}
\nonumber
G_{o o}&=&\int_0^\infty \!\!\!du\,P(u)
\left(
\begin{array}{cc}
    1&-F(u)\Sigma\\
    0&1-F(u)^2\Sigma
\end{array}
\right)^{-1}
\left(
\begin{array}{cc}
    1&F(u)\\
    F(u)&F(u)^2
\end{array}
\right)\\
&=&\int_0^\infty \!\!\!du\,P(u)
\frac{1}{1-F(u)^2\Sigma}
\left(
\begin{array}{cc}
    1&F(u)\\
    F(u)&F(u)^2
\end{array}
\right).\label{eq:Goo2x2itoG0Sigma}
\end{eqnarray}
The local propagator $G_{o o}$ can be obtained from the impurity action in terms of the Weiss field $\mathcal{G}_0$, 
which is now a $2\times2$ matrix, by Grassmann integration
(\ref{eq:effectivetheory}), and reads:
\begin{align}
&G_{o o}\nonumber
=\frac{\int\mathcal{D}[\bar\phi_o,\phi_o]\,\left(
\begin{array}{cc}
    \phi_{o\sigma 0\rho}\bar\phi_{o\sigma 0\rho} & \phi_{o\sigma 0\rho}\bar\phi_{o\sigma 1\rho}\\
    \phi_{o\sigma 1\rho}\bar\phi_{o\sigma 0\rho} & \phi_{o\sigma 1\rho}\bar\phi_{o\sigma 1\rho}
\end{array}
\right)
e^{-\bar\phi_o \mathcal{G}_0^{-1} \phi_o}}{\int\mathcal{D}[\bar\phi_o,\phi_o]\,(1+\prod_{\rho,\sigma}\phi_{o\sigma\rho}\bar\phi_{o\sigma\rho})e^{-\bar\phi_o \mathcal{G}_0^{-1}\phi_o}}\nonumber\\
&=\frac{\mathcal{G}_0+\left([\mathcal{G}_0]_{11}\right)^3{\rm Det}\mathcal{G}_0
\left(
\begin{array}{cc}
1&0\\
0&0    
\end{array}
\right)}
{1+\left([\mathcal{G}_0]_{11}\right)^4}.\label{eq:G2x2itoK}
\end{align}
We also write out the self-consistency condition Eq.~(\ref{eq:selfcon1}) in component form,
using the ``self-energy''~(\ref{eq:Sigma2x2}):
\begin{eqnarray}
G_{o o}&=&(\mathcal{G}_0^{-1}-\Sigma)^{-1}\nonumber\\
&=&\frac{\mathcal{G}_0-{\rm Det}\mathcal{G}_0\Sigma
\left(
\begin{array}{cc}
1&0\\
0&0    
\end{array}
\right)}
{1-[\mathcal{G}_0]_{11}\Sigma}.\label{eq:G2x2selfcon}
\end{eqnarray}
The $\alpha=1$ component $[\mathcal{G}_0]_{11}$ is the scalar Weiss field $\mathcal{G}_0$ in
Sect.\,\ref{sec:givendelta}, since equating (\ref{eq:G2x2itoK}) and (\ref{eq:G2x2selfcon}) gives
\begin{equation}
\Sigma=-\left([\mathcal{G}_0]_{11}\right)^{3}.
\end{equation}
Then equating (\ref{eq:Goo2x2itoG0Sigma}) and (\ref{eq:G2x2itoK}) gives three self-consistency
conditions:
\begin{align}
&\frac{[\mathcal{G}_0]_{11}}{1+\left([\mathcal{G}_0]_{11}\right)^4}=\int_0^\infty du\,P(u)\frac{F(u)^2}{1+F(u)^2 [\mathcal{G}_0]_{11}^{3}},\nonumber\\
&[\mathcal{G}_0]_{01}=[\mathcal{G}_0]_{10}=\frac{[\mathcal{G}_0]_{11}\int_0^\infty du\,P(u)\frac{F(u)}{1+F(u)^2 [\mathcal{G}_0]_{11}^{3}}}{\int_0^\infty du\,P(u)\frac{F(u)^2}{1+F(u)^2 [\mathcal{G}_0]_{11}^{3}}},\nonumber\\
&[\mathcal{G}_0]_{00}=\frac{[\mathcal{G}_0]_{11}\int_0^\infty du\,P(u)\frac{1}{1+F(u)^2 [\mathcal{G}_0]_{11}^{3}}}{\int_0^\infty du\,P(u)\frac{F(u)^2}{1+F(u)^2 [\mathcal{G}_0]_{11}^{3}}}.\label{eq:K2x2sol}
\end{align}
We have already encountered the first line above in Sect.\,\ref{sec:givendelta}. It allows us to find $[\mathcal{G}_0]_{11}$ (previously simply called $\mathcal{G}_0$) for given $F(u)$.
Subsequently, the next two lines then give the remaining components of the matrix $\mathcal{G}_0$ in terms of this solution.

The kinetic energy per unit volume is calculated from knowledge of the momentum distribution $n_{\bm k}$ (\ref{eq:nk})
\begin{align}
&\frac{E_\text{kin}}{V}=2\int_0^{2\pi}\prod_{a=1}^d \frac{dk_a}{2\pi}\en_{\bm k}n_{\bm k}\nonumber\\
&=\frac{\left|E_{\text{kin},0}\right|}{V}\left[1-4\sqrt\pi \int_0^\infty du\,\frac{P(u)u}{1+f(u)^2}\right],\label{eq:kin}
\end{align}
where
\begin{equation}
\frac{\left|E_{\text{kin},0}\right|}{V}=\frac{2t^*}{\sqrt{\pi}},\label{eq:kin0}
\end{equation}
and as defined previously
\begin{equation}
f(u)=\left([\mathcal{G}_0]_{11}\right)^{3/2}F(u).\label{eq:f2x2}
\end{equation}

In order to calculate the double-occupancy, we note that the local electronic charge is
\begin{equation}
n_{o\sigma}=\left(\psi_{o\sigma 0-}+\psi_{o\sigma 0+}^\dagger\right)\left(\psi_{o\sigma 0-}^\dagger+\psi_{o\sigma 0+}\right).\label{eq:nitopsi}
\end{equation}
Anti-normal ordering the double-occupancy, and omitting contributions that vanish due to particle-hole symmetry, we obtain 
\begin{align}
&\left<n_{o\uparrow}n_{o\downarrow}\right>\nonumber\\
&=\frac{1}{4}+\left<\psi_{o\uparrow 0+}^\dagger \psi_{o\downarrow 0+}^\dagger \psi_{o\downarrow 0-}^\dagger \psi_{o\uparrow 0-}^\dagger\right>+\text{c.c.}\label{eq:double1}
\end{align}
We compute the expectation value by Grassmann integration (\ref{eq:effectivetheory}), and use Eq.~(\ref{eq:K2x2sol})
to express the result in terms of $f(u)$ and $[\mathcal{G}_0]_{11}$ only.
\begin{align}
&\left<\psi_{o\uparrow 0+}^\dagger \psi_{o\downarrow 0+}^\dagger \psi_{o\downarrow 0-}^\dagger \psi_{o\uparrow 0-}^\dagger\right>\nonumber\\
&=\frac{-\left([\mathcal{G}_0]_{01}\right)^4}{1+\left([\mathcal{G}_0]_{11}\right)^4}\nonumber\\
&=-\left[\int_0^\infty du\,P(u)\frac{F(u)}{1+F(u)^2 [\mathcal{G}_0]_{11}^{3}}\right]^4\left[1+\left([\mathcal{G}_0]_{11}\right)^4\right]^3\nonumber\\
&=-\left[\int_0^\infty du\,P(u)\frac{f(u)}{1+f(u)^2}\right]^4\frac{\left[1+\left([\mathcal{G}_0]_{11}\right)^4\right]^3}{\left([\mathcal{G}_0]_{11}\right)^{6}}.\label{eq:double2}
\end{align}
Putting Eq.~(\ref{eq:kin}) and Eq.~(\ref{eq:double2}) together, we find the following expression 
for the total energy $\left<H\right> = E_\text{kin} + U \left<n_{o\uparrow}n_{o\downarrow}\right>$:
\begin{equation}
\frac{\left<H\right>}{\left|E_{\text{kin},0}\right|}=1-4\sqrt\pi \int_0^\infty du\,\frac{P(u)u}{1+f(u)^2}\nonumber\\
+2\bar U\left(1-8R^3 W^4\right),
\end{equation}
where
\begin{equation}
R=\frac{1+\left([\mathcal{G}_0]_{11}\right)^4}{\left([\mathcal{G}_0]_{11}\right)^2},
\qquad 
W =\int_0^\infty du\, P(u)\,\frac{f(u)}{1+f(u)^2},
\label{eq:defR}
\end{equation}
and we have defined the dimensionless interaction:
\begin{equation}
\bar U=\frac{VU}{8|E_{\text{kin},0}|}=\frac{\sqrt{\pi}U}{16t^*}.\label{eq:defbarU}
\end{equation}
Eliminating Eq.~(\ref{eq:Kiton}), we get 
$([\mathcal{G}_0]_{11})^4=c/(1-c)$ with the density per spin of particle-hole excitations:
\begin{equation}
c = \int_0^\infty du\, P(u) n(u) =
\int_0^\infty du\, P(u)\,\frac{f(u)^2}{1+f(u)^2},
\end{equation}
which is now a function of $f(u)$ only.
The factor $R$ reduces to $R=[c(1-c)]^{-1/2}$, and 
the energy becomes an unconstrained functional of $f(u)$.
Minimization gives:
\begin{eqnarray}
&& f(u)^2+2a(1+u/b)f(u)-1=0,\qquad \\
&& \Rightarrow f(u)=\sqrt{a^2\left(1+\frac{u}{b}\right)^2+1}-a\left(1+\frac{u}{b}\right),\label{eq:fitoab}
\end{eqnarray}
with $a=\frac38\frac{W(1-2c)}{c(1-c)}$ and $b=\frac{6\bar U W^4(1-2c)}{\sqrt\pi\,[c(1-c)]^{5/2}}.$
Injecting Eq.~(\ref{eq:fitoab}) in the expression for $W$ and $c$, one can directly solve for $a$ and $b$.
Let us now analyze these equations analytically in the limit of weak or strong interactions.

\subsection{Weak coupling}
In the limit $\bar U\to 0$, $1/b$ becomes large, so that
\begin{equation}
f(u)\simeq \frac{f_0}{1+\frac{u}{b}},\label{eq:fsmall}
\end{equation}
with $f_0$ small.
We expand to order $f_0^2$. All integrals over $u$ can be done analytically.
To shorten notation, we define
\begin{eqnarray}
y(b)&=&e^{-b^2}\left[\pi\text{Erfi}(b)-\text{Ei}(b^2)\right],\nonumber\\
z(b)&=&b^2y(b)-\sqrt{\pi}b+1,
\end{eqnarray}
with $\text{Erfi}(x)=2\int_0^x dt\,e^{t^2}/\sqrt{\pi}$ and $\text{Ei}(x)=\mathcal P\int_{-\infty}^x dt\,e^t/t$.
In terms of these functions, we obtain 
\begin{equation}
([\mathcal{G}_0]_{11})^4 = \frac{c}{1-c} \simeq\frac{f_0^2 b}{\sqrt{\pi}}z(b).\label{eq:solK11}
\end{equation}
Substituting this into Eq.~(\ref{eq:defR}), we find
\begin{equation}
R\simeq([\mathcal{G}_0]_{11})^{-2}=\sqrt{\frac{\sqrt{\pi}}{bz(b)}}f_0^{-1}.
\end{equation}
The expectation value of the energy, expressed in terms of $f_0$ and $b$ reads
\begin{align}
\frac{\left<H\right>}{\left|E_{\text{kin},0}\right|}=&-1+4b^2\left[\frac{y(b)}{2}-z(b)\right]f_0^2\nonumber\\
&+2\bar U\left[1-\frac{b^{5/2}y(b)^4}{2\pi^{5/4}z(b)^{3/2}}f_0\right].\label{eq:ensmall}
\end{align}
We minimize this with respect to $f_0$, and obtain
\begin{align}
&f_0=\frac{b^{1/2}y(b)^4}{8\pi^{5/4}z(b)^{3/2}\left[\frac{y(b)}{2}-z(b)\right]}\bar U\nonumber,\\    
&\frac{\left<H\right>}{\left|E_{\text{kin},0}\right|}=-1+2\bar U -\frac{b^3y(b)^8}{16\pi^{5/2}z(b)^{3}\left[\frac{y(b)}{2}-z(b)\right]}\bar U^2.\label{eq:enitob}
\end{align}
The final minimization with respect to $b$ can also be done analytically. Using $y'(b)=-2z(b)/b$, which follows from the definitions of $y$ and $z$, the minimum is found where $8z(b)=3y(b)$, i.e.,
\begin{equation}
y(b)=\frac{8\left(\sqrt{\pi}\,b-1\right)}{8b^2-3},\label{eq:bopt}
\end{equation}
which yields $b=1.2005$ and $y(b)=1.0578$.
Finally, we evaluate the amount $Z$ by which the momentum distribution jumps at
the Fermi surface, and the energy, at the minimum.
Since $z=3y/8$ there, both take a simple closed form:
\begin{align}
&Z=1-2f_0^2=1-\frac{1024}{27}\frac{b\,y(b)^3}{\pi^{5/2}}\bar U^2=1-3.0806\, \bar U^2\nonumber\\
&\frac{\left<H\right>}{\left|E_{\text{kin},0}\right|}=-1+2\,\bar U-\frac{256}{27}\frac{b^3y(b)^4}{\pi^{5/2}}\bar U^2\nonumber\\
&\qquad\qquad\ \ =-1+2\,\bar U-1.1741\,\bar U^2.\label{eq:smallsol}
\end{align}
Since $Z$ has been determined after finding the optimal trial state for a Hamiltonian known to host a Fermi liquid,
it is appropriate to refer to it here as the quasiparticle weight, although we did not obtain it as the residue of the quasiparticle pole
of the dynamical single-particle Green function.

The exact ground state energy of the Hubbard model in $d\to\infty$, obtained from second-order
perturbation theory~\cite{Metzner1989b,Muller-Hartmann1989b}, reads:

\begin{eqnarray}
\nonumber
\frac{E^{(2)}}{|E_{\text{kin},0}|}
&=& -1+2\,\bar U -\frac{U^2}{2t^*|E_0|}\int_{u_i>0}\frac{\prod_{i=1}^4 P(u_i)\,du_i}{u_1+u_2+u_3+u_4},\\
\nonumber
&=& -1+2\,\bar U -\frac{8}{\sqrt\pi}\bar U^2\int_0^\infty \mathrm{erfcx}(x)^4dx,\\
&=&-1+2\,\bar U -1.1905\,\bar U^2.\label{eq:exacten}
\end{eqnarray}
where we used $1/(u_1+\ldots)=\int_0^\infty e^{-\lambda (u_1+\ldots)}d\lambda$, and the Laplace transform of $P(u)$.
Furthermore $\text{erfcx}(x)=2 e^{x^2}\int_x^{\infty} dt\,e^{-t^2}/\sqrt{\pi}$.
From Eq.~(\ref{eq:smallsol}), the single-flavor ansatz accounts for $1.1741/1.1905=98.6\%$ of 
the second-order in interaction energy contribution to
the ground state energy. The exact expansion of the quasiparticle weight reads similarly:
\begin{equation}
Z=1-\frac{64}{\pi}\bar U^2\int_0^\infty x\,\mathrm{erfcx}(x)^3dx=
1-4.687\,\bar U^2 
\end{equation}
so that the trial state $\left|\Psi\right>$ accounts for $66\%$ of the second order contribution.

In contrast, the Gutzwiller state (\ref{eq:Gutzwiller}) gives~\cite{Brinkman1970,Vollhardt1984,Metzner1989b}
\begin{eqnarray}
\frac{E_G}{|E_{\text{kin},0}|} &=& (-1+2\bar U - \bar U^2) = -(1-\bar U)^2,\nonumber\\
Z&=&1-\bar U^2\label{eq:Eg}
\end{eqnarray}
accounting for $84\%$ of the second order contribution to the energy and $22\%$ of the second order contribution to the quasiparticle weight.

\begin{figure}[h]
    \begin{center}
    \includegraphics[width=.99\columnwidth]{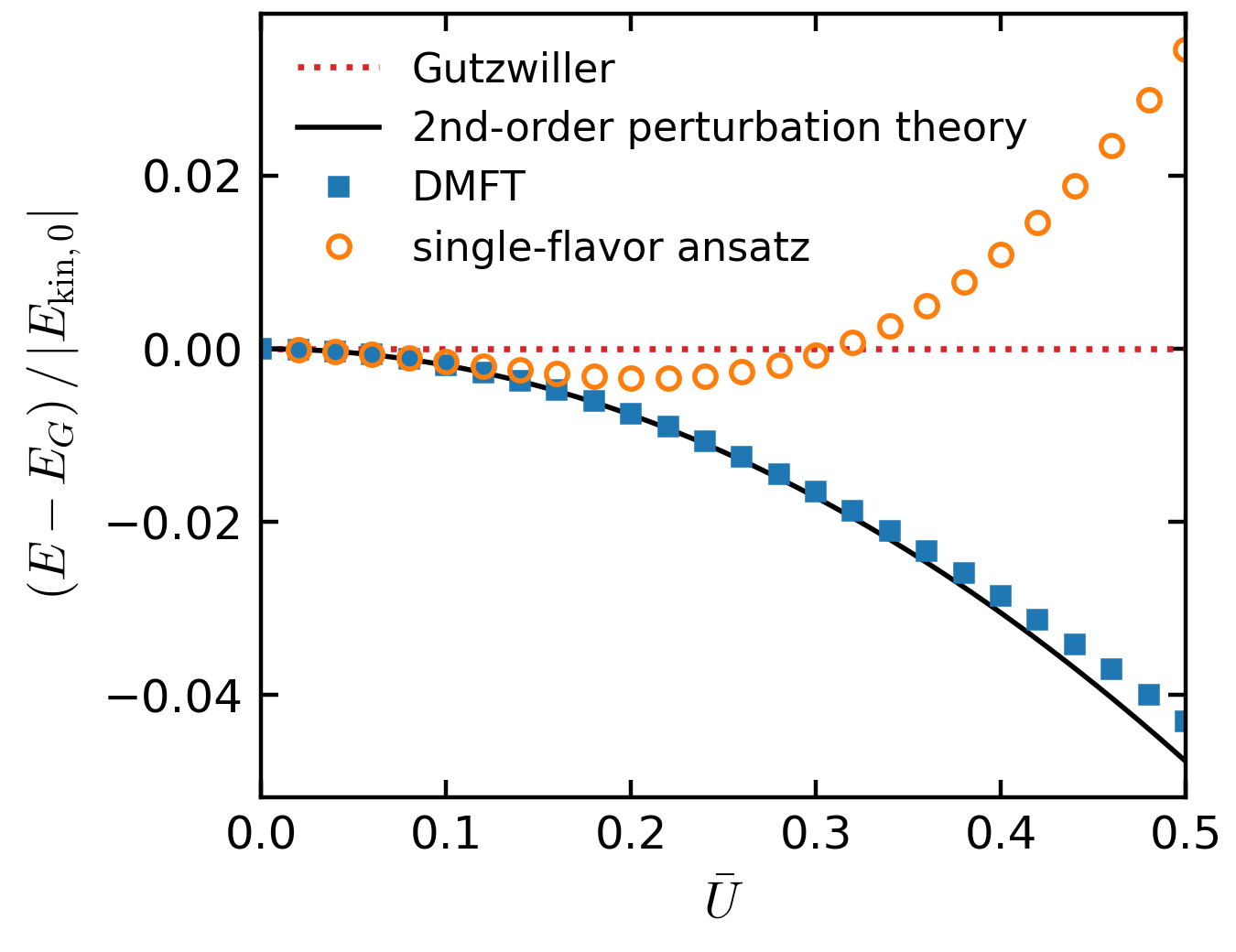}
    \end{center}
    \caption{{\bf Circles:} Difference between the optimal variational energy of the single flavor Ansatz 
    and the energy of the Gutzwiller state (\ref{eq:Eg}), in units of the non-interacting ground state energy, 
    as a function of dimensionless interaction strength $\bar U$ (\ref{eq:defbarU}), for weak coupling. 
    {\bf Squares:} DMFT calculation for the Hubbard model using EDIpack~\cite{EDIpack}.
    {\bf Solid line:} Second order perturbation theory.\label{fig:energy}}
\end{figure}

In Figure\,\ref{fig:energy}, we compare the optimal variational energy of the single flavor ansatz for the Hubbard model to the
energy of the Gutzwiller state, as a function of the dimensionless interaction strength $\bar U$. The Gutzwiller 
ansatz predicts a Mott transition at $\bar U=1$. However, between $\bar U=0$ and $\bar U=0.31$, the single flavor ansatz
has a lower energy, owing to a more accurate description of the Fermi liquid phase. Nonetheless, the single flavor ansatz
deviates significantly from the perturbative energy for $\bar U\gtrsim 0.1$. DMFT calculations
for the Hubbard model on the hypercubic lattice were performed using EDIpack~\cite{EDIpack}, and  show that in fact, 
the perturbative energy is relatively accurate for $\bar U\lesssim 0.3$. In other words, the single flavor ansatz 
has a $\bar U^4$ contribution that is significantly larger than the true fourth-order correction to the ground state energy,
a familiar failure encountered when some processes are summed to all orders while others, responsible for large cancellations, are omitted.

\subsection{Strong coupling}
In the limit $\bar U\to \infty$, the optimal envelope $F(u)$ becomes independent of energy, so that
\begin{equation}
f(u)=f_\infty.\label{eq:finf}
\end{equation}
Since this limit is dominated by interaction energy, this indicates the intuitively obvious fact that a flat envelope 
minimizes the double occupancy. 
The pertinent question to ask is how much suppression of the double-occupancy is possible.
For a flat envelope function, the double-occupancy reads
\begin{equation}
\left<n_{o\uparrow}n_{o\downarrow}\right>=\frac{1}{4}-\frac{1}{\left(2+f_\infty^2\right)^{3/2}}\frac{f_\infty}{1+f_\infty^2}.\label{eq:doubleflat}
\end{equation}
We minimize with respect to $f_\infty$, and find the optimal value $f_\infty^2=(\sqrt{3}-1)/2$.
For the optimal $f_\infty$, the quasiparticle weight and double-occupancy read
\begin{align}
&Z=\frac{1-f_\infty^2}{1+f_\infty^2}=\frac{3-\sqrt{3}}{\sqrt{3}+1}\simeq0.4641,\nonumber\\
&\left<n_{o\uparrow}n_{o\downarrow}\right>=0.1283.
\end{align}
One can show for any envelope function, not necessarily the one in Eq.~(\ref{eq:fitoab}), that this bound on 
double occupancy derives from the inequality $W\le\frac12\sqrt{2c(1-2c)}$, and is indeed saturated 
by the flat envelope.
Thus any single flavor coupled cluster double state $\left|\Psi\right>$ is only expressive enough 
to reduce the $1/4$ double-occupancy of the uncorrelated state by about $50\%$.
It is therefore not able to capture the Mott transition.
This is consistent with double-occupation results obtained for Hubbard chains in Ref.\,\cite{Ai2026} 
with a coupled cluster method that is to some extent robust against the singularities of the conventional 
evaluation technique, suggesting that what is seen in Ref.\,\cite{Ai2026} is probably diagnostic 
of the limitations of the coupled cluster doubles state.

\section{Beyond single flavor doubles}
\label{sec:beyonddoubles}

\subsection{Multi-flavor doubles}
The coupled cluster method employed in quantum chemistry assumes a far more general cluster operator
that the single-flavor double amplitude that we have employed thus far~\cite{BartlettMusial2007,Shavitt2009}.
Before concluding that the limited suppression of double-occupancy is a general feature of coupled cluster doubles states, we should explore the 
effect of incorporating more flavors.

Mostly closed form expressions for large parts of the calculation, can
still be obtained for two flavors. To incorporate more than two  flavors,
one could instead map the effective single-site Grassmann integral back onto a fermionic few-body
problem with the number of fermions equal to four times the number of flavors, which could then be
solved numerically. This approach, along with the question of multi-flavor envelope choice, and the
subsequent numerical minimization strategy in a larger parameter space, are beyond the scope of the
current work. We therefore  limit ourselves here to the case of two flavors. 
The rank-two cluster operator reads
\begin{align}
T_j^\dagger=&\ \sum_{\alpha\beta\gamma\delta} A_{\alpha\beta\gamma\delta}\psi_{j\uparrow\alpha+}^\dagger \psi_{j\downarrow\beta+}^\dagger \psi_{j\downarrow\gamma-}^\dagger \psi_{j\uparrow\delta-}^\dagger\nonumber\\
&+B \sum_{\alpha\beta\gamma\delta} 
\sum_{\sigma\in\{\uparrow,\downarrow\}} \psi_{j\sigma 1+}^\dagger \psi_{j\sigma 2+}^\dagger
\psi_{j\sigma 2-}^\dagger \psi_{j\sigma 1-}^\dagger.
\end{align}
The conditions $
A_{\alpha\beta\gamma\delta}=A_{\beta\alpha\delta\gamma}=A_{\delta\gamma\beta\alpha}$
with the flavor indices $\alpha,\beta,\gamma,\delta\in\{1,2\}$ encode spin-rotation and particle-hole symmetry. This leaves
seven independent components of the $A$-tensor.

We define the following notation:
\begin{equation}
\tilde1=2,~\tilde 2=1,~\varepsilon_{1,2}=-\varepsilon_{2,1}=1.
\end{equation}
Grassmann integration (\ref{eq:effectivetheory}) then gives the local ``impurity'' propagator:
\begin{align}
&G_{o\alpha,o\beta}=\frac{N_{\alpha\beta}}{D},\nonumber\\
&N_{\alpha\alpha}=\left[1+B^2 \left({\rm Det}\mathcal{G}_0\right)^2\right][\mathcal{G}_0]_{\alpha\alpha}\nonumber\\
&~~~~~~~~~+{\rm Det}\mathcal{G}_0 A_{\tilde \alpha\beta\gamma\delta} A_{\tilde \alpha\beta'\gamma'\delta'} [\mathcal{G}_0]_{\beta\beta'}[\mathcal{G}_0]_{\gamma\gamma'}[\mathcal{G}_0]_{\delta\delta'}\nonumber\\
&N_{\alpha\tilde\alpha}=\left[1+B^2 \left({\rm Det}\mathcal{G}_0\right)^2\right][\mathcal{G}_0]_{\alpha\tilde \alpha}\nonumber\\
&~~~~~~~~~-{\rm Det}\mathcal{G}_0 A_{\tilde \alpha\beta\gamma\delta} A_{ \alpha\beta'\gamma'\delta'} [\mathcal{G}_0]_{\beta\beta'}[\mathcal{G}_0]_{\gamma\gamma'}[\mathcal{G}_0]_{\delta\delta'}\nonumber\\
&D=1+ A_{\alpha\beta\gamma\delta} A_{\alpha'\beta'\gamma'\delta'} [\mathcal{G}_0]_{\alpha\alpha'}[\mathcal{G}_0]_{\beta\beta'}[\mathcal{G}_0]_{\gamma\gamma'}[\mathcal{G}_0]_{\delta\delta'}\nonumber\\
&~~~~~+2B^2  \left({\rm Det}\mathcal{G}_0\right)^2\nonumber\\
&+\frac{1}{4}\left(A_{\alpha\beta\gamma\delta}
A_{\tilde\alpha\tilde\beta\tilde\gamma\tilde\delta}\varepsilon_{\alpha\tilde\alpha}\varepsilon_{\beta\tilde\beta}\varepsilon_{\gamma\tilde
\gamma}\varepsilon_{\delta\tilde\delta}+2B^2\right)^2 \left({\rm Det}\mathcal{G}_0\right)^4,
\label{eq:GitoK}
\end{align}
with Einstein summations implied on dummy indices.

We are interested to see if it is possible or not to improve the suppression of double-occupancy 
within the general family of coupled cluster doubles (recall that for any single envelope states it
is bounded from below by 0.1283, and from above by the maximum value 1/4 for free fermions). We therefore forget about kinetic energy
which penalizes single-particle excitations far from the Fermi energy. Given that double-occupancy is equally
sensitive to single-particle excitations at all momenta, we pick the envelopes that define the fermion flavors
to be flat in a certain energy window. We define $v=\text{erf}(u)$ so that $\int_0^\infty du\, P(u)=\frac{1}{2} \int_0^1 dv$.
Then we define energy windows each consisting of half the particle- or hole orbitals
\begin{align}
&\mathcal P_{2-}=\left\{\en|\text{erf}(\en/2t^*)\in(-1,-1/2)\right\},\nonumber\\
&\mathcal P_{1-}=\left\{\en|\text{erf}(\en/2t^*)\in(-1/2,0)\right\},\nonumber\\
&\mathcal P_{1+}=\left\{\en|\text{erf}(\en/2t^*)\in(0,1/2)\right\},\nonumber\\
&\mathcal P_{2+}=\left\{\en|\text{erf}(\en/2t^*)\in(1/2,1)\right\}.
\end{align}
The flavor index $\alpha$ distinguishes which half of the particle or hole orbitals are being addressed.
\begin{eqnarray}
\psi_{j\sigma\alpha+}&=&\frac{1}{\sqrt{V}}\sum_{\varepsilon_{\bm k}\in \mathcal P_{\alpha+}} e^{i\bm k\cdot \bm r_j}c_{\bm k\sigma},\nonumber\\
\psi_{j\sigma\alpha-}&=&\frac{1}{\sqrt{V}}\sum_{\varepsilon_{\bm k}\in \mathcal P_{\alpha-} }e^{-i\bm k\cdot \bm r_j}c_{\bm k\sigma}^\dagger.\label{eq:2flat}
\end{eqnarray}
For this specific choice of flat envelopes, the analog Dyson equation (\ref{eq:dyson}) reduces to
\begin{equation}
G_{o\alpha,o\beta}=\frac{1}{4}\frac{\delta_{\alpha\beta}}{1-\Sigma_{\alpha\alpha}}.\label{eq:eqGflat}
\end{equation}
Equating this to the self-consistency condition Eq.~(\ref{eq:selfcon1})
then leads to an expression for the elements of $\mathcal{G}_0^{-1}$ in terms of the elements of the ``self-energy''.
\begin{equation}
[\mathcal{G}_0^{-1}]_{\alpha\alpha}=4-3\Sigma_{\alpha\alpha},~[\mathcal{G}_0^{-1}]_{\alpha\tilde\alpha}=\Sigma_{\alpha\tilde\alpha}. \label{eq:KitoSigma}
\end{equation}     
It is important to note that while $G_{o o}$ in Eq.~(\ref{eq:eqGflat}) is diagonal in flavor, and only 
depends on the diagonal components of $\Sigma$, neither $\mathcal{G}_0^{-1}$ nor $\Sigma$ is diagonal in flavor. Also
only when Eq.~(\ref{eq:GitoK}), Eq.~(\ref{eq:eqGflat}), and Eq.~(\ref{eq:KitoSigma}) are
simultaneously satisfied, will the off-diagonal elements $N_{12}=N_{21}$ in Eq.~(\ref{eq:GitoK})
evaluate to zero. If Eq.~(\ref{eq:KitoSigma}), with an arbitrary choice of $\Sigma$ 
is substituted into Eq.~(\ref{eq:GitoK}) instead, in general $N_{12}=N_{21}\ne0$ will result.
As a consistency check of the above equations, we note that if we choose all $A_{\alpha\beta\gamma\delta}=A$, and $B=0$,
we obtain a solution in which all elements of $\mathcal{G}_0^{-1}$ are equal, and the equations reduce to what one would get
for a single flavor $\psi_{m\sigma\rho}=\psi_{m\sigma 1\rho}+\psi_{m\sigma 2\rho}$.

We now set up an iterative loop, that starts with an arbitrary guess for $\Sigma$ and converges to a simultaneous
solution of $G$, $\mathcal{G}_0^{-1}$ and $\Sigma$ that solve Eq.~(\ref{eq:GitoK}), Eq.~(\ref{eq:eqGflat}) and Eq.~ (\ref{eq:KitoSigma}).
It works as follows
\begin{enumerate}
\item Guess an initial $\Sigma_0$.
\item Iterate the following steps until $|\Sigma_n-\Sigma_{n-1}|$ is sufficiently small.
\begin{enumerate}
\item Compute $\mathcal{G}_{0,n}^{-1}$ corresponding to $\Sigma_n$ from Eq.~(\ref{eq:KitoSigma}).
\item Compute $\left. G_{o o} \right._n$ corresponding to $\mathcal{G}_{0,n}^{-1}$ from Eq.~(\ref{eq:GitoK}).
\item Compute an updated guess $\Sigma'$ for the ``self-energy'', from $\Sigma'=\mathcal{G}_{0,n}^{-1}-[\left.G_{o o}\right._n]^{-1}$.
\item Pick the next ``self-energy'' as a mixture of the previous one and the updated guess: $\Sigma_{n+1}=\alpha \Sigma'+(1-\alpha)\Sigma_n$.
\end{enumerate}
\end{enumerate}

The above results allow us to find the matrix $\mathcal{G}_0^{-1}$ for given tensor $A$ and scalar $B$
that define a two-flavor doubles coupled cluster state $\left|\Psi\right>$. 
In Appendix \ref{app:doubocc2} we derive an expression for the double-occupation in terms of $\mathcal{G}_0^{-1}$. 
We use it to perform simulated annealing searches for the minimum 
double-occupancy $\left<n_{o\uparrow}n_{o\downarrow}\right>$ 
with respect to the seven independent coefficients in the tensor $A$ and the coefficient $B$
that together define $\left|\Psi\right>$. 
Inspection suggests that the double-occupancy landscape in this parameter space  
is free of metastable local minima or barren plateaus. 
The minimum that each search finds 
has all $A$ coefficients equal and $B=0$, so that the state reduces to 
the same single-flavor state (\ref{eq:finf}) we found previously, with double-occupancy $0.1283$.
The fact that we find no improvement {\em whatsoever} when allowing two flavors, strongly suggests that
a single flavor with a flat envelope yields the lowest double occupancy within the full family
of multi-flavor doubles, although this has not been proven.

\subsection{Triples and more}

Other than increasing the number of flavors, the family of considered states can be generalized
by adding terms to the cluster operator $T_j^\dagger$ (\ref{eq:clusterop}) that create three (triples), or more particle-hole pairs~\cite{BartlettMusial2007,Shavitt2009},
in addition to the already included ones that create two pairs (doubles). In Ref.\,\cite{Paldus1984} 
it was found that the inclusion of quadruples within the traditional coupled cluster method
improved the quality of results significantly for Hubbard chains at strong coupling. When triples, etc. are included,
operators $\psi_{m\sigma\alpha-}^\dagger \psi_{n\sigma\beta+}^\dagger$ describing the excitation of a single particle of spin $\sigma$ across the
Fermi surface may now acquire a non-zero expectation value, due to the cross-term of a double and
triple excitation, (and many other higher-order cross terms).
In Appendix \ref{app:C} we generalize the formalism of the main text to handle this situation.

The inclusion of triples adds a sum of terms of the form:
\begin{equation}
    C_{\alpha\beta\gamma\alpha'\beta'\gamma'}\psi_{j\sigma\alpha+}^\dagger \psi_{j\uparrow\beta+}^\dagger \psi_{j\downarrow\gamma+}^\dagger
    \psi_{j\downarrow\gamma'-}^\dagger \psi_{j\uparrow\beta'-}^\dagger \psi_{j\sigma\alpha'-}^\dagger,
\end{equation}
to the cluster operator $T^\dagger_j$ (\ref{eq:clusterop}). Here, summation over all repeated Greek flavor and spin indices is understood.
We now specialize to the case of two flavors, which significantly constrains the number of possible triples amplitudes
in the cluster operator. We assume particle-hole and spin rotation symmetry, which constrains the triples part of the cluster operators 
to a single independent amplitude.
Furthermore, there is also only a single quadruples amplitude, involving the product of all eight $\psi_{\sigma\alpha\rho}^\dagger$ operators.
In Appendix \ref{app:3and4} we work out all the equations necessary for calculating the double occupancy for this trial state. 

We repeat the previous analysis, i.e. we minimize the double occupancy in the case of two
flavors with the flat profiles of Eq.~(\ref{eq:2flat}). 
Again inspection suggests a benign double occupancy landscape in the variational parameter space which is now ten-dimensional.  
Simulated annealing still finds that within this enlarged family
the single ``flat" flavor state with double occupancy $0.1283$ is the lowest possibility.
Unlike in the case of multi-flavor doubles, we do not think this settles the expressivity question
for triples, quadruples, etc. In principle, we think that the inclusion of triples and quadruples, with more than two flavors, and with
arbitrary profiles, should at least be able to achieve variational results comparable in accuracy 
to the next order $\bar U^4$ in perturbation theory. As mentioned before,
in the context of a relatively short Hubbard chain, quadruples lead to improved behavior at strong correlation, with
the caveat that evaluation of expectation values were subject to conventional coupled cluster approximations \cite{Paldus1984}.
It would therefore be surprising if 
using triples and quadruples with a sufficient number of additional flavors cannot lower double-occupancy 
somewhat below the best possible value achievable with doubles.
We ascribe the failure of our triples and quadruples calculation to improve the double-occupancy result 
in the two-flavor case
to the fact that it introduces only two independent new variational parameters. 
Regarding the inclusion of triples and higher order terms, we therefore conclude with an open problem:
To explore the expressivity of higher than doubles coupled cluster states in the infinite dimensional limit of the Hubbard model,
requires further work on the optimal choice for the structure and number of flavors. 

\section{Conclusions}
\label{sec:conclusions}

We have shown that within a class of coupled cluster states for fermions on the hypercubic lattice in the limit dimension $d\to\infty$, 
the evaluation of certain expectation values maps onto a static few-body problem associated with a single lattice site.
Arbitrary local observables (such as the double occupancy) and all single-particle quantities
(such as the single-particle the momentum distribution function) can be obtained by the formalism. The same power counting 
that underlies DMFT~\cite{Metzner1989b,Muller-Hartmann1989,Georges1996} reduces the influence of all 
other sites on a given site to a Gaussian kernel $\mathcal{G}_0^{-1}$, and closes the problem through the cavity 
relation $G_{o o}=(\mathcal{G}_0^{-1}-\Sigma)^{-1}$, with $\Sigma$ a site-diagonal analog of the DMFT
self-energy that represents irreducible corrections to the static one-particle amplitudes.
Interestingly, in contrast to DMFT, the resulting `impurity problem' involves only a finite number 
of Grassmann variables, and can be solved by elementary integration.
For the simplest member of the class, a Fermi sea
dressed by the exponential of a rank-one (single fermion `flavor') doubles excitations, 
we achieved an exact closed-form evaluation. 

For a single flavor, 
any particle-hole symmetric $n(\en)$ can be realized, including distributions that are completely smooth
at the Fermi energy. By the gossamer metal construction~\cite{Kollar2003}, each such state is the
unique ground state of an explicit, positive semi-definite parent Hamiltonian with correlated
hopping and a gapless single-particle tunnelling spectrum.

Applied to the paramagnetic half-filled Hubbard model, the ansatz performs well at weak coupling,
because energy-dependent flavor envelopes approximate well the energy denominators of perturbation theory.
(This is just rank-one Tucker factorization of the cluster amplitude.)
A two-parameter closed-form envelope recovers $98.6\%$ of the exact second-order energy and $66\%$ of
the exact second-order suppression of the quasiparticle weight, compared with $84\%$ and $22\%$ for
the Gutzwiller state. At strong coupling, the same family of coupled cluster states fails: 
the double occupancy cannot be lowered below $0.1283$, and neither extending to more flavors,
nor to the triples and quadruples lower this value. We attribute this to a structural difference 
between the two kinds of state. The Gutzwiller factor in Eq.~(\ref{eq:Gutzwiller}) is a function 
of the idempotent operator $n_{j\uparrow}n_{j\downarrow}$, and is linear in it, so that it can fully
project out doubly occupied sites. The exponential of a local excitation operator built from
particle and hole creation operators mimics the Gutzwiller factor at small amplitude, but lacks this
projecting property. This connects with the well-documented difficulty of coupled cluster doubles
theory with strong or static correlation~\cite{BartlettMusial2007,Bulik2015,Qiu2017}. The failures
of truncated coupled cluster theory for finite Hubbard
systems~\cite{Paldus1984,Stein2014,Bulik2015,Ai2026} thus acquire an exact analogue in which no
approximation in the evaluation is involved, albeit for a sub-class of the states that depend on
single-particle momenta only through $\en_{\bm k}$. We further emphasize that only the single
energy-envelope case was explored exhaustively. 

Nonetheless, this last point illustrates what we regard as the main value of the infinite-dimensional setting: it is a laboratory in which the quality of a coupled cluster ansatz can be separated cleanly from the quality of the scheme used to evaluate it. In finite dimensions, a deficient wave function and a truncated, non-variational amplitude equation cannot easily be disentangled. It should further be noted that in order to explore the coupled cluster state's ability to capture Mott physics, we have intentionally demanded translational invariance. Thus we prevented the state from breaking sublattice symmetry and developing anti-ferromagnetic Néel order -- the true ground state of the hyper-cubic Hubbard model at large $U$. The correct statement is therefore ``the single-flavor coupled cluster double state provides a poor description of a Mott parent state of strongly correlated fermions'' and not that it``provides a poor description of the large $U$ hyper-cubic Hubbard model in the $d\to\infty$ limit''. Preliminary calculations that we have performed, indicate that, as expected, coupled clusters states capture Néel order well when sublattice symmetry breaking, that cuts off strong correlations, is allowed.  

Several extensions suggest themselves. While development of competitive variational ground states
\cite{Lanata2017,Cheng2021,Cheng2021b} for correlated lattice fermions in dimension $d\to\infty$ was not our main 
motivation here, our states might yet contribute to future developments in this area. Their niche might be problems where a knob is required that directly controls 
the occupation of single-particle energy orbitals. Conventional coupled cluster results for Hubbard chains \cite{Paldus1984} suggest        
the inclusion of more flavors (and more sophisticated envelope choices) in the triples
and quadruples calculation may enhance expressivity in the right direction. If the family of states we consider can be integrated into VDAT \cite{Cheng2021,Cheng2021b} despite their use of not-completely-local
two-body operators, this would allow exact evaluation with Gutzwiller-dressed coupled cluster states $P_G e^{-T^\dagger}\left|F\right>$, where $T^\dagger$ controls Fermi-liquid behavior and $P_G$ controls local charge fluctuations.    
Finally, it would be interesting to know whether or not the parent Hamiltonian of a state with smooth $n(\en)$ and gapless single-particle tunnelling spectrum retains some trace
of a Fermi surface in other observables or response functions.

\appendix
\section{Cavity method for coupled clusters}
\label{app:A}
Here we derive the equation $G_{o o}=(\mathcal{G}_0^{-1}-\Sigma)^{-1}$ that relates the local Green
function on site $o$ to the Weiss field $\mathcal{G}_0$, using only basic linear algebra identities,
the scaling  $d^{-|\bm r_{m}-\bm r_{n}|/2}$ of $G_0$ and $G_0^{-1}$, and the locality
of $\Sigma$.
The key physical observation is that in $d\to\infty$ the cavity carries the same
local ``self-energy'' as the full lattice.

\emph{Cavity matrices.} 
We start by recalling a well-known relation between the inverse of a sub-block and the inverse of a full matrix, and 
presenting it in notation that is convenient for the site/cavity partitioning. 
The coupled cluster theory involves matrices with a combination of spatial, spin, flavor and
particle/hole indices. The crux of the cavity mapping lies in the influence
functional~(\ref{eq:influence}), where inverse ``propagators'' are truncated to sites $m,n\neq o$.
This motivates us to define, for an invertible lattice matrix $M$ (akin to a propagator), its cavity
version:
\begin{equation}
M^{[o]}\equiv\Big(\big[M^{-1}\big]_{m,n\ne o}\Big)^{-1}\label{eq:cavdef}
\end{equation}
that is obtained by inverting $M$, deleting the row and the column of site $o$, and inverting back within 
the cavity (keeping only sites $m,n\ne o$). 
Here and below, matrix elements such as $M_{mn}$ denote the block of $M$ with site indices $(m,n)$,
which is itself a matrix in the spin, flavor and particle/hole indices. Products such as
$\sum_l M_{ml}N_{ln}$ are block products, with the internal indices summed inside each product.
Let us first show that the cavity matrix can be related 
to the original matrix by:
\begin{equation}
M^{[o]}_{mn}=M_{mn}-M_{mo}\,[M_{oo}]^{-1}M_{on},\label{eq:blockinv}
\end{equation}
for $m,n\ne o$. From $\sum_l [M^{-1}]_{ml}M_{ln}=\delta_{mn}$, where~$\delta_{mn}$ 
includes the unit matrix of internal indices, we split the sum:
$\sum_{l\neq o}[M^{-1}]_{ml}M_{ln} + [M^{-1}]_{mo}M_{on} =\delta_{mn}$.
We get
$\sum_{l\neq o}[M^{-1}]_{ml}M_{lo}+ [M^{-1}]_{mo}M_{oo} =0$
by setting $n=o$ and $m\neq o$, thus 
$[M^{-1}]_{mo}=-\sum_{l\ne o}[M^{-1}]_{ml}M_{lo}[M_{oo}]^{-1}$
by inversion. We stress here that the site-local quantity $M_{oo}$ is still a matrix 
in spin, flavor, particle/hole indices.
Inserting this result into the previous split sum, we get:
$\sum_{l\ne o}[M^{-1}]_{ml}\big(M_{ln}-M_{lo}[M_{oo}]^{-1}M_{on}\big)=\delta_{mn}$.
For $m,n\ne o$, the expression under parenthesis is the inverse of $[M^{-1}]_{m,n\ne o}$, i.e.\ $M^{[o]}$,
which proves (\ref{eq:blockinv}).
We will derive also another useful result:
\begin{equation}
\big[M_{oo}\big]^{-1}=[M^{-1}]_{oo}-\sum_{m,n\ne o}[M^{-1}]_{om}\,M^{[o]}_{mn}\,[M^{-1}]_{no},
\label{eq:schur}
\end{equation}
that relates $[M_{oo}]^{-1}$, the inverse of the local block $M_{oo}$, to
$[M^{-1}]_{oo}$, the local block of the inverse. To prove the wanted result, we start 
again with $M^{-1}M=1$, taken at $m\ne o$ and $n=o$:
$\sum_{l\ne o}[M^{-1}]_{ml}M_{lo}=-[M^{-1}]_{mo}M_{oo}$.
Multiplying this from the left by $M^{[o]}$, which by definition~(\ref{eq:cavdef}) is the inverse of $[M^{-1}]_{m,n\ne o}$, 
gives for $l\ne o$: $M_{lo}=-\sum_{m\ne o}M^{[o]}_{lm}[M^{-1}]_{mo}M_{oo}$. Inserting this into
equation $M^{-1}M=1$ with $m=n=o$, namely $[M^{-1}]_{oo}M_{oo}+\sum_{l\ne o}[M^{-1}]_{ol}M_{lo}=1$, 
and multiplying by $[M_{oo}]^{-1}$, one finds then (\ref{eq:schur}).

\emph{Cavity propagators.} 
The Grassmann influence functional~(\ref{eq:influence}) has Gaussian weight
$e^{-\sum_{m,n\ne o}\bar\phi_m[G_0^{-1}]_{mn}\phi_n}$.
The associated bare propagator for the diagrammatics in the cavity is thus the inverse 
of $[G_0^{-1}]_{m,n\ne o}$, which by definition~(\ref{eq:cavdef}) is exactly $G_0^{[o]}$.
By (\ref{eq:blockinv}), the bare propagator of the cavity is therefore the one of the full 
lattice minus the propagators running through site $o$:
\begin{equation}
[G_0^{[o]}]_{mn}=[G_0]_{mn}-[G_0]_{mo}[G_0]_{oo}^{-1}[G_0]_{on}.
\label{eq:cavprop}
\end{equation}
We saw, by invoking the linked-cluster expansion (\ref{eq:linkedcluster}), that the full influence
functional reduces in $d\to\infty$ to a Gaussian part, denoted:
\begin{equation}
\mathcal I[\phi_o,\bar \phi_o]=\exp[\bar\phi_o([G_0^{-1}]_{o o}-\mathcal{G}_0^{-1})\phi_o],\label{eq:infquadaap}
\end{equation}
with
\begin{equation}
[G_0^{-1}]_{o o}-\mathcal{G}_0^{-1}=\sum_{m,n\ne o}[G_0^{-1}]_{o m}\,\big[G^{(c)}_{[o]}\big]_{m n}\,[G_0^{-1}]_{n o},\label{eq:kernel}
\end{equation}
where $G^{(c)}_{[o]}$ is the connected ``interacting'' propagator of the cavity. By ``interacting'',
we mean here that the influence functional~(\ref{eq:influence}) contains the non-quadratic terms
$T_j + \bar T_j$.
We now come to the 
central 
step of the proof, namely that in $d\to\infty$:
\begin{equation}
G^{(c)}_{[o]}=G^{[o]},\qquad\text{i.e.}\qquad \big[G^{(c)}_{[o]}\big]^{-1}=\big[G^{-1}\big]_{m,n\ne o},
\label{eq:cavityG}
\end{equation}
namely the ``interacting'' propagator of the cavity is the cavity version (\ref{eq:cavdef}) of 
the ``interacting'' propagator on the full lattice.
The cavity is a lattice theory of the same type as the full lattice, except with site $o$ removed.
It uses bare propagator $G_0^{[o]}$, so its full propagator obeys a Dyson equation 
$[G^{(c)}_{[o]}]^{-1}=[G_0^{-1}]_{m,n\ne o}-\Sigma_{[o]}$. Ths cavity ``self-energy'' 
$\Sigma_{[o]}$ is diagonal in site indices, by the counting below Eq.~(\ref{eq:dyson}), 
but it is {\it a priori} not spatially uniform.
The diagrams for $[\Sigma_{[o]}]_{jj}$ differ from those for $\Sigma_{jj}$ only by the
absence of vertices on site $o$ and by the use of the cavity bare propagator $G_0^{[o]}$ from
Eq.~(\ref{eq:cavprop}), which receive the subtraction of lattice bare propagators via site $o$. 

Any difference between the self-energy on the full lattice and the one on the cavity
must be diagrams with one point pinned
on site $o\ne j$: an internal vertex in $\Sigma_{jj}$, or in $\left[\Sigma_{[o]}\right]_{jj}$ a point through which a $G_0$ propagator line is forced to pass
because of the relation (\ref{eq:cavprop}). 
An irreducible diagram cannot be disconnected by cutting a single line, and inserting a point on a line via Eq.~\ref{eq:cavprop}
preserves this, so the pinned point is joined to $j$ by at least two independent paths (sharing no
propagator line). Holding it fixed on site $o$ while all other sums are performed gives a
contribution of order at most $d^{-\sum_a|r_{oa}-r_{ja}|}$ where $r_{ja}$ is the $a$-component of vector $\bm r_j$ , and a diagram of finite order has
finitely many such points. Hence $[\Sigma_{[o]}]_{jj}=\Sigma+\mathcal
O(d^{-\sum_a|r_{oa}-r_{ja}|})$. These corrections enter (\ref{eq:kernel}) through propagation from
$o$ to $j$ and back, $\sim d^{-\sum_a|r_{oa}-r_{ja}|}$, so with $\sim d^L$ sites at Manhattan
distance $L$ their total is $\sum_{L\ge1}d^{L}\,d^{-L}\,d^{-L}=\mathcal O(d^{-1})$. Up to terms that
vanish as $d\to\infty$, we thus obtain
$[G^{(c)}_{[o]}]^{-1}=[G_0^{-1}-\Sigma]_{m,n\ne o}=[G^{-1}]_{m,n\ne o}$, which is (\ref{eq:cavityG}). 
Thus removing site $o$ does not change the ``self-energy'' of the other sites in the $d\to\infty$ limit. 
From here on, the cavity Green function may simply be written $G^{[o]}$.

\emph{Final result.} Let us first apply (\ref{eq:schur}) to $M=G$. Locality of $\Sigma$ gives $[G^{-1}]_{om}=[G_0^{-1}]_{om}$ and
$[G^{-1}]_{no}=[G_0^{-1}]_{no}$ for $m,n\ne o$, and $[G^{-1}]_{oo}=[G_0^{-1}]_{oo}-\Sigma$. 
With (\ref{eq:cavityG}) and (\ref{eq:kernel}), the resulting (\ref{eq:schur}) reads:
\begin{align}
[G_{oo}]^{-1}&=[G_0^{-1}]_{oo}-\Sigma-\sum_{m,n\ne o}[G_0^{-1}]_{om}\,G^{[o]}_{mn}\,[G_0^{-1}]_{no}\nonumber\\
&=\mathcal{G}_0^{-1}-\Sigma,
\label{eq:finalcav}
\end{align}
which is the desired result~(\ref{eq:selfcon1}).
Applying instead (\ref{eq:blockinv}) to $M=G$ turns (\ref{eq:finalcav}) into
\begin{align}
&[G_0^{-1}]_{o o}-\mathcal{G}_0^{-1}\nonumber\\
&=\sum_{m,n\ne o}[G_0^{-1}]_{o m}\left(G_{m n}-G_{m o}[G_{oo}]^{-1}G_{o n}\right)[G_0^{-1}]_{n o},\label{eq:cavity1}
\end{align}
which is the standard expression of the cavity Green function in DMFT~\cite{Georges1996}, namely
the full $G_{mn}$ minus the contribution of paths that visit site $o$. 

\section{Double occupancy of doubles with two flavors}
\label{app:doubocc2} 

Here we work out expressions for the double occupancy in the case of two flavors with flat envelopes, as introduced in Eq.~\ref{eq:2flat}.
We can split position-basis operators for site $o$ as
$c_{o \sigma}=p_\sigma+h_\sigma^\dagger$, where $p_\sigma=\psi_{o\sigma 1+}+\psi_{o\sigma 2+}$ and
$h_\sigma=\psi_{o\sigma 1-}+\psi_{o\sigma 2-}$.
Keeping in mind spin rotation symmetry, particle-hole symmetry, and the fact that the number of
particles plus holes is a multiple of four, we can write double occupancy as
\begin{align}
&\left<n_{o\uparrow}n_{o\downarrow}\right>=\left<(p_\uparrow ^\dagger+h_\uparrow)(p_\uparrow+h_\uparrow^\dagger)(p_\downarrow ^\dagger+h_\downarrow)(p_\downarrow+h_\downarrow^\dagger)\right>\nonumber\\
=&\frac{1}{4}+\underbrace{\left<p_\uparrow^\dagger h_\uparrow^\dagger p_\downarrow^\dagger h_\downarrow^\dagger\right>+\text{cc}}_{=T_1/D}+\underbrace{\left<p_\downarrow h_\downarrow h_\uparrow^\dagger p_\uparrow^\dagger\right> + \text{cc}}_{T_2/D}\nonumber\\
&+\underbrace{2\left<p_\uparrow p_\uparrow^\dagger p_\downarrow p_\downarrow^\dagger\right>}_{T_3/D}+\underbrace{(-2)\left<p_\uparrow h_\downarrow h_\downarrow^\dagger p_\uparrow^\dagger \right>}_{T_4/D}
\end{align}
Using Grassmann integration (\ref{eq:effectivetheory}), we find the following expressions for $T_1$ to $T_4$, with sums over repeated indices implied:
\begin{widetext}
\begin{eqnarray}
T_1&=&-2\sum_{\alpha_2\beta_2\gamma_2\delta_2}A_{\alpha_1\beta_1\gamma_1\delta_1} [\mathcal{G}_0]_{\alpha_1\alpha_2}[\mathcal{G}_0]_{\beta_1\beta_2}[\mathcal{G}_0]_{\gamma_1\gamma_2}[\mathcal{G}_0]_{\delta_1\delta_2}\nonumber\\
&&-\left(A_{\alpha_1\beta_1\gamma_1\delta_1} A_{\tilde\alpha_1\tilde\beta_1\tilde\gamma_1\tilde\delta_1}\varepsilon_{\alpha_1\tilde\alpha_1}\varepsilon_{\beta_1\tilde\beta_1}\varepsilon_{\gamma_1\tilde \gamma_1}\varepsilon_{\delta_1\tilde\delta_1}+2B^2\right)
A_{\tilde\alpha_2\tilde\beta_2\tilde\gamma_2\tilde\delta_2}\varepsilon_{\alpha_2\tilde\alpha_2}\varepsilon_{\beta_2\tilde\beta_2}\varepsilon_{\gamma_2\tilde \gamma_2}\varepsilon_{\delta_2\tilde\delta_2}\left({\rm Det}\mathcal{G}_0\right)^4,\nonumber\\
T_2&=&4B\sum_{\alpha_2\delta_2}A_{\alpha_1\beta_1\gamma_1\delta_1}\varepsilon_{\beta_1\tilde\beta_1}\varepsilon_{\gamma_1\tilde\gamma_1}[\mathcal{G}_0]_{\alpha_1\alpha_2}[\mathcal{G}_0]_{\delta_1\delta_2}\left({\rm Det}\mathcal{G}_0\right)^2,\nonumber\\
T_3&=&2\left(\sum_{\alpha_1\alpha_2}[\mathcal{G}_0]_{\alpha_1\alpha_2}\right)^2+2 A_{\tilde \alpha_1\tilde \beta_1\gamma_2\delta_2} A_{\tilde\delta_1\tilde\gamma_1\beta_2\alpha_2}
\varepsilon_{\alpha_1\tilde\alpha_1}\varepsilon_{\beta_1\tilde\beta_1}\varepsilon_{\gamma_1\tilde \gamma_1}\varepsilon_{\delta_1\tilde\delta_1}[\mathcal{G}_0]_{\delta_2\alpha_2}[\mathcal{G}_0]_{\gamma_2\beta_2}\left({\rm Det}\mathcal{G}_0\right)^2,\nonumber\\
T_4&=&-2\left(\sum_{\alpha_1\alpha_2}[\mathcal{G}_0]_{\alpha_1\alpha_2}\right)^2
-2 A_{\tilde \alpha_1\beta_2\tilde \gamma_1\delta_2} A_{\tilde\delta_1\gamma_2\tilde \beta_1\alpha_2}
\varepsilon_{\alpha_1\tilde\alpha_1}\varepsilon_{\beta_1\tilde\beta_1}\varepsilon_{\gamma_1\tilde \gamma_1}\varepsilon_{\delta_1\tilde\delta_1}[\mathcal{G}_0]_{\alpha_2\delta_2}[\mathcal{G}_0]_{\gamma_2\beta_2}\left({\rm Det}\mathcal{G}_0\right)^2.
\end{eqnarray}
Note that
\begin{equation}
T_3+T_4=2 \left(A_{\tilde \alpha_1\tilde \beta_1\gamma_2\delta_2} A_{\tilde\delta_1\tilde\gamma_1\beta_2\alpha_2}-A_{\tilde \alpha_1\beta_2\tilde \gamma_1\delta_2} A_{\tilde\delta_1\gamma_2\tilde \beta_1\alpha_2}\right)
\varepsilon_{\alpha_1\tilde\alpha_1}\varepsilon_{\beta_1\tilde\beta_1}\varepsilon_{\gamma_1\tilde \gamma_1}\varepsilon_{\delta_1\tilde\delta_1}[\mathcal{G}_0]_{\delta_2\alpha_2}[\mathcal{G}_0]_{\gamma_2\beta_2}\left({\rm Det}\mathcal{G}_0\right)^2.
\end{equation}
\end{widetext}

\section{Triples, etc. and non-zero single-particle-hole propagators.}
\label{app:C}

When incorporating triples, etc. into the cavity method, the cavity Green function acquires anomalous components corresponding to  
$\psi_{m\sigma\alpha-}^\dagger \psi_{n\sigma\beta+}^\dagger$ propagators etc. Only after self-consistency is imposed, does momentum
conservation assert itself, to ensure that $\psi_{o\sigma\alpha-}^\dagger \psi_{o\sigma\beta+}^\dagger$ expectation values are zero.
The inclusion of triples, etc. in the cluster operator therefore requires a generalization of the 
formalism used in the main text, in a way that is formally similar to the Green function formalism in the presence of singlet Cooper pairing.  
Arranging the different types of normal-ordered single-particle additive operators into a matrix and signifying this additional matrix structure with
a check overscript, we define
\begin{equation}
\check{G}_{m\alpha,n\beta}=\left(\begin{array}{cc}\left<\psi_{m\sigma\alpha+}\psi_{n\sigma\beta+}^\dagger\right>&\left<\psi_{m\sigma\alpha+}\psi_{n\sigma\beta-}\right>\\
    \left<\psi_{m\sigma\alpha-}^\dagger \psi_{n\sigma\beta+}^\dagger\right>&\left<\psi_{n\sigma\beta-} \psi_{m\sigma\alpha-}^\dagger\right>\end{array}\right).\label{eq:nambuG}
\end{equation}
Note the reversal of the order of indices in the lower left entry.
Organizing the diagrams that make up $\check{G}_{m\alpha,n\beta}$ into a geometric series of ``self-energy'' contributions connected by
zero-order propagator lines, we identify four different types of self-energies, that are naturally arranged into a $2\times 2$ matrix.
\begin{equation}
\check{\Sigma}_{\alpha m,\beta n}=\delta_{m,n}\left(\begin{array}{cc}\Sigma^{(+)}_{\alpha\beta}&S_{\alpha\beta}\\\bar S_{\alpha\beta}& \Sigma^{(-)}_{\beta \alpha}\end{array}\right).\label{eq:nambuSigma}
\end{equation}
(Proportionality to identity in site indices is again due to the scaling of propagators in $d\to\infty$.)
Note again the reversal of the order of indices in the lower left entry.
In $\Sigma^{(+)}_{\alpha\beta}$ and $\Sigma^{(-)}_{\beta\alpha}$, the left subscript refers to a flavor of the 
incoming propagator line that was amputated, and the right subscript refers to a flavor of the 
outgoing propagator line that was amputated. 
In $S_{\alpha\beta}$ both subscripts refer to flavor indices of outgoing propagator lines, while in $\bar S_{\beta\alpha}$ both subscripts refer to flavor indices of incoming propagator lines.
Treated as matrices in particle-hole, flavor and site indices, $\check{G}$, $\check{G}_0$ and $\check{\Sigma}$ satisfy the Dyson equation
\begin{equation}
\check{G}=\left(1-\check{G}_0\check{\Sigma}\right)^{-1}\check{G}_0.\label{eq:nambuDyson}
\end{equation}
The Gaussian effective action with kernel $\mathcal{G}_0^{-1}$ now generalizes to
\begin{equation}
S_o=\sum_{\sigma\alpha\beta}\left(\bar \phi_{\sigma\alpha} \bar \varphi_{\sigma\alpha}\right)\left(\begin{array}{cc} [\mathcal{G}_0^{(+)}]^{-1}_{\alpha\beta} & L_{\alpha\beta}\\
    \bar L_{\alpha\beta}&-[\mathcal{G}_0^{(-)}]^{-1}_{\beta\alpha}\end{array}\right)\left(\begin{array}{c}\phi_{\sigma\beta}\\ \varphi_{\sigma\beta}\end{array}\right),\label{eq:nambuK}
\end{equation}
where $\bar\phi_{\sigma\alpha}=\bar\phi_{o\sigma\alpha+}$, $\bar \varphi_{\sigma\alpha}=\phi_{o\sigma\alpha-}$, $\phi_{\sigma\alpha}=\phi_{o\sigma\alpha+}$, and $\varphi_{\sigma\alpha}=\bar\phi_{o\sigma\alpha-}$.
We treat the entries of $[\mathcal{G}_0^{(+)}]^{-1}$, $[\mathcal{G}_0^{(-)}]^{-1}$, $L$ and $\bar L$ as independent parameters, thus allowing us to compute the generating function 
\begin{equation}
Z={\int\mathcal{D}[\bar\phi,\phi, \bar \varphi, \varphi]\,e^{-T_o-\bar T_o-S_o}}.\label{eq:genfun}
\end{equation}
The self-consistency condition $G_{o o}=(\mathcal{G}_0^{-1}-\Sigma)^{-1}$ derived in Appendix\,\ref{app:A} generalizes to
\begin{align}
&\left(\begin{array}{cc}\left<\psi_{o\sigma\alpha+}\psi_{o\sigma\beta+}^\dagger\right>&\left<\psi_{o\sigma\alpha+}\psi_{o\sigma\beta-}\right>\\
    \left<\psi_{o\sigma\alpha-}^\dagger \psi_{o\sigma\beta+}^\dagger\right>&-\left<\psi_{o\sigma\beta-} \psi_{o\sigma\alpha-}^\dagger\right>\end{array}\right)\nonumber\\
    &=\left[\left(\begin{array}{cc} [\mathcal{G}_0^{(+)}]^{-1}-\Sigma^{(+)} & L-S\\
    \bar L - \bar S&-\left([\mathcal{G}_0^{(-)}]^{-1}-\Sigma^{(-)}\right)^T\end{array}\right)^{-1}\right]_{\alpha\beta}.\label{eq:nambuSC}
\end{align}
Note the additional minus sign, appearing in the lower right components of Eq.~(\ref{eq:nambuK}) and Eq.~(\ref{eq:nambuSC}) as compared to Eq.~\ref{eq:nambuG}.
After taking the desired derivatives of $Z$, time-reversal symmetry allows us to set 
\begin{equation}
\bar S=S^T, \quad \bar L=L^T,
\end{equation}
while particle-hole symmetry allows us to set
\begin{align}
    &[\mathcal{G}_0^{(+)}]^{-1}=[\mathcal{G}_0^{(-)}]^{-1}=\mathcal{G}_0^{-1}, \quad \Sigma^{(+)}=\Sigma^{(-)}=\Sigma,\nonumber\\ 
    &L^T=-L, \quad S^T=-S.
\end{align}

\section{Two flavor triples and quadruples}
\label{app:3and4}

Here we detail the inclusion of triples and quadruples within the two-flavor family of states.  
We write the contribution of $\exp(-\sum_j T_j^\dagger)$ to the Grassmann integrand on site $o$ as    
\begin{align}
&e^{-\bar T_o}=\underbrace{1}_1\nonumber\\
&\underbrace{-A_{\alpha\beta\gamma\delta}\bar \phi_{\uparrow\alpha}\bar \phi_{\downarrow\beta}\varphi_{\downarrow\gamma}\varphi_{\uparrow\delta}-B \bar\phi_{\sigma 1}\bar \phi_{\sigma 2}\varphi_{\sigma 2}\varphi_{\sigma 1}}_4\nonumber\\
&\underbrace{-C\en_{\alpha\tilde\alpha}\bar \phi_{\uparrow\tilde \alpha}\bar \phi_{\downarrow\tilde \alpha} \bar \phi_{\sigma\alpha}\varphi_{\sigma\tilde\alpha}\varphi_{\downarrow\alpha}\varphi_{\uparrow\alpha}}_6\nonumber\\
&\underbrace{+D\bar\phi_{\uparrow 1}\bar \phi_{\downarrow 1}\bar \phi_{\uparrow 2}\bar \phi_{\downarrow 2}\varphi_{\uparrow 1}\varphi_{\downarrow 1}\varphi_{\uparrow 2}\varphi_{\downarrow 2}}_8.\label{eq:et}
\end{align}
Here we write the coefficient in front of the term with eight Grassmann numbers as $D$ and treat it as an independent parameter.
This amounts to including not only triples, but also quadruples, in the theory restricted to two flavors.

For two flavors, a brute-force algebraic calculation of $Z$ leads to a closed-form expression and symbolic algebra software
can subsequently be used to take derivatives of $Z$. For this purpose, we write $Z$ as
\begin{align}
Z=\Big[&\left(1|8\right)+\left(1|6\right)+\left(1|4\right)+1\nonumber\\
+&\left(8|1\right)+\left(6|1\right)+\left(4|1\right)\nonumber\\
+&\left(4|8\right)+\left(4|6\right)+\left(4|4\right)\nonumber\\
+&\left(8|4\right)+\left(6|4\right)\nonumber\\
+&\left(6|8\right)+\left(6|6\right)\nonumber\\
+&\left(8|6\right)\nonumber\\
+&\left(8|8\right)\Big]\det M_\uparrow \det M_\downarrow,\label{eq:Zsum}
\end{align}
with numbers refering to products of terms with labels as in Eq.~\ref{eq:et}.
Here we have defined a matrix
\begin{equation}
M_\sigma=\left(\begin{array}{cccc}
    M^{(\sigma)}_{1+,1+}&M^{(\sigma)}_{1+,2+}&M^{(\sigma)}_{1+,1-}&M^{(\sigma)}_{1+,2-}\\
    M^{(\sigma)}_{2+,1+}&M^{(\sigma)}_{2+,2+}&M^{(\sigma)}_{2+,1-}&M^{(\sigma)}_{2+,2-}\\
    M^{(\sigma)}_{1-,1+}&M^{(\sigma)}_{1-,2+}&-M^{(\sigma)}_{1-,1-}&-M^{(\sigma)}_{1-,2-}\\
    M^{(\sigma)}_{2-,1+}&M^{(\sigma)}_{2-,2+}&-M^{(\sigma)}_{2-,1-}&-M^{(\sigma)}_{2-,2-}\end{array}\right), 
\end{equation}
corresponding to an arbitrary Gaussian kernel for the action $S_o$ in Eq.~\ref{eq:nambuK}.
After differentiating the generating functional to extract expectation values, one makes the substitution
\begin{equation}
M_\sigma\to M_o=\left(\begin{array}{cccc}
    [\mathcal{G}_0^{-1}]_{11}&[\mathcal{G}_0^{-1}]_{12}&0&L\\
    {}[\mathcal{G}_0^{-1}]_{12}&[\mathcal{G}_0^{-1}]_{22}&-L&0\\
    0&-L&-[\mathcal{G}_0^{-1}]_{11}&-[\mathcal{G}_0^{-1}]_{12}\\
    L&0&-[\mathcal{G}_0^{-1}]_{12}&-[\mathcal{G}_0^{-1}]_{22}\end{array}\right).
\end{equation}
The matrix $M_o$ has inverse
\begin{align}
M_o^{-1}=&\frac{1}{\det\mathcal{G}_0^{-1}+L^2}\nonumber\\
&\times\left(\begin{array}{cccc}
    [\mathcal{G}_0^{-1}]_{22}&-[\mathcal{G}_0^{-1}]_{12}&0&L\\
    -[\mathcal{G}_0^{-1}]_{12}&[\mathcal{G}_0^{-1}]_{11}&-L&0\\
    0&-L&-[\mathcal{G}_0^{-1}]_{22}&[\mathcal{G}_0^{-1}]_{12}\\
    L&0&[\mathcal{G}_0^{-1}]_{12}&-[\mathcal{G}_0^{-1}]_{11}\end{array}\right),
\end{align}
and determinant 
\begin{equation}
\det M_o=\left(\det\mathcal{G}_0^{-1}+L^2\right)^2.
\end{equation}

We label rows and columns of the matrix $M_\sigma$ by the pairs $(\alpha \rho)$, with  $\alpha=1,2$ and $\rho=\pm$. We define
\begin{align}
Q_\sigma(\alpha_1\rho_1,\ldots,\alpha_n\rho_n|\alpha_1'\rho_1',\ldots,\alpha_n'\rho_n')
\end{align}
as the determinant of the $n\times n$ submatrix of $M_\sigma^{-1}$ with rows $(\alpha_1\rho_1),\ldots,(\alpha_n\rho_n)$ and columns $(\alpha_1'\rho_1'),\ldots,(\alpha_n'\rho_n')$.
With the aid of these definitions, we have
\begin{equation}
\left(1|8\right)=DQ_\uparrow(1-,2-|1+,2+)Q_\downarrow(1-,2-|1+,2+),
\end{equation}
\begin{align}
\left(1|6\right)=&-C\en_{\alpha,\tilde \alpha}\Big[Q_\uparrow(1-,2-|1+,2+)Q_\downarrow(\alpha-|\tilde \alpha+)\nonumber\\
&+Q_\downarrow(1-,2-|1+,2+)Q_\uparrow(\alpha-|\tilde \alpha+)\Big],
\end{align}
\begin{align}
\left(1|4\right)=&-A_{\alpha\beta\gamma\delta}Q_\downarrow(\gamma-|\beta+)Q_\uparrow(\delta-|\alpha+)\nonumber\\
&-B\left[Q_\uparrow(1-,2-|1+,2+)+Q_\downarrow(1-,2-|1+,2+)\right],
\end{align}
\begin{align}
\left(4|8\right)=&-A_{\alpha\beta\gamma\delta}DQ_\downarrow(1-,2-,\beta+|1+,2+,\gamma-)\nonumber\\
&\times Q_\uparrow(1-,2-,\alpha+|1+,2+,\delta-),
\end{align}
\begin{align}
\left(4|6\right)=&C\en_{\alpha,\tilde \alpha}\Big\{A_{\alpha'\beta'\gamma'\delta'}\big[Q_\downarrow(1-,2-,\beta'+|1+,2+,\gamma'-)\nonumber\\
&~~~~~~~~~~~~~~~~~~~~~~~~~\times Q_\uparrow(\alpha-,\alpha'+|\tilde \alpha+,\delta'-)\nonumber\\
&~~~~~~~~~~~~~~~~~~~+Q_\uparrow(1-,2-,\alpha'+|1+,2+,\delta'-)\nonumber\\
&~~~~~~~~~~~~~~~~~~~~~~~~~\times Q_\downarrow(\alpha-,\beta'+|\tilde \alpha+,\gamma'-)\big]\nonumber\\
&+B\big[\det M_\downarrow^{-1} Q_\uparrow(\alpha-|\tilde \alpha+)\\
&~~~~~+\det M_\uparrow^{-1} Q_\downarrow(\alpha-|\tilde \alpha+)\nonumber\\
&~~~~~+Q_\downarrow(1-,2-|1+,2+)\nonumber\\
&~~~~~~~~~\times Q_\uparrow(1+,2+,\alpha-|1-,2-,\tilde \alpha+)\nonumber\\ 
&~~~~~+Q_\uparrow(1-,2-|1+,2+)\nonumber\\
&~~~~~~~~~\times Q_\downarrow(1+,2+,\alpha-|1-,2-,\tilde \alpha+)
\big]\Big\},
\end{align}
\begin{align}
\left(4|4\right)=&A_{\alpha\beta\gamma\delta}A_{\alpha'\beta'\gamma'\delta'}\nonumber\\
&\times Q_\downarrow(\gamma-,\beta'+|\gamma'-,\beta)Q_\uparrow(\delta-,\alpha'+|\delta'-,\alpha)\nonumber\\
&+BA_{\alpha\beta\gamma\delta}\big[Q_\uparrow(1+,2+,\delta-|1-,2-,\alpha+)\nonumber\\
&~~~~~~~~~~~~~~~~~~~~~~~~~~~\times Q_\downarrow(\gamma-|\beta+)\nonumber\\
&~~~~~~~~+Q_\uparrow(1-,2-,\alpha+|1+,2+,\delta-)\nonumber\\
&~~~~~~~~~~~~~~~~~~~~~~~~~~~\times Q_\downarrow(\beta+|\gamma-)\nonumber\\
&~~~~~~~~+Q_\downarrow(1+,2+,\gamma-|1-,2-,\beta+)\nonumber\\
&~~~~~~~~~~~~~~~~~~~~~~~~~~~\times Q_\uparrow(\delta-|\alpha+)\nonumber\\
&~~~~~~~~+Q_\downarrow(1-,2-,\beta+|1+,2+,\gamma-)\nonumber\\
&~~~~~~~~~~~~~~~~~~~~~~~~~~~\times Q_\uparrow(\alpha+|\delta-)\big]\nonumber\\
&+B^2\big[\det M_\downarrow^{-1}+\det M_\uparrow^{-1}\nonumber\\
&+Q_\downarrow(1-,2-|1+,2+)Q_\uparrow(1+,2+|1-,2-)\nonumber\\
&+Q_\downarrow(1+,2+|1-,2-)Q_\uparrow(1-,2-|1+,2+)\big],\\
\left(6|8\right)=&-C\en_{\alpha,\tilde \alpha} D\big[\det M_\downarrow^{-1}Q_\uparrow(1-,2-,\tilde\alpha+|1+,2+,\alpha-)\nonumber\\
&+\det M_\uparrow^{-1}Q_\downarrow(1-,2-,\tilde\alpha+|1+,2+,\alpha-)\big].
\end{align} 
\begin{align}
\left(6|6\right)=&-C^2\en_{\alpha,\tilde \alpha}\en_{\alpha',\tilde \alpha'}\big[\det M_\downarrow^{-1}Q_\uparrow(\alpha-,\tilde\alpha'+|\alpha'-,\tilde\alpha+)\nonumber\\
&~~~-Q_\downarrow(1-,2-,\tilde\alpha'+|1+,2+,\alpha'-)\nonumber\\
&~~~~~~~~~~\times Q_\uparrow(1+,2+,\alpha-,|1+,2+,\tilde\alpha+)\nonumber\\
&+\det M_\uparrow^{-1}Q_\downarrow(\alpha-,\tilde\alpha'+|\alpha'-,\tilde\alpha+)\nonumber\\
&~~~-Q_\uparrow(1-,2-,\tilde\alpha'+|1+,2+,\alpha'-)\nonumber\\
&~~~~~~~~~~\times Q_\downarrow(1+,2+,\alpha-,|1+,2+,\tilde\alpha+)\big],\\
\left(8|8\right)&=D^2\det M^{-1}_\uparrow \det M^{-1}_\downarrow.
\end{align}
The remaining terms in Eq.~(\ref{eq:Zsum}) are related to those already calculated by the transformation
\begin{align}
&\left(x|y\right)\to\left(x|y\right)\implies\nonumber\\
&Q_\sigma(\alpha_1\rho_1,\ldots,\alpha_n\rho_n|\alpha_1'\rho_1',\ldots,\alpha_n'\rho_n')\to\nonumber\\
&Q_\sigma(\rho_1'\alpha_1',\ldots,\rho_n'\alpha_n'|\rho_1\alpha_1,\ldots,\rho_n\alpha_n).
\end{align}
We firstly use $Z$ to obtain an expression for $\check G$ in terms of $\mathcal{G}_0^{-1}$ and $L$, via
\begin{equation}
\check G_{o\alpha\rho,o\alpha'\rho'}=\left.\frac{1}{Z}\frac{\partial Z}{\partial M^{(\sigma)}_{\alpha'\rho',\alpha\rho}}\right|_{M_\sigma=M_o}.\label{eq:tripcon1}
\end{equation}
We want to see what the lowest achievable double-occupancy is, and so specialize to the previously considered case of two
flavors with piece-wise flat profiles (\ref{eq:2flat}). In this situation, momentum conservation prohibits nonzero
$\check G_{o\alpha\rho,o\alpha'\tilde\rho}$, and $\check G$ is simply related to $\Sigma$ via
\begin{equation}
\check G_{o\alpha\rho,o\alpha'\rho'}=\frac{1}{4}\frac{\delta_{\rho\rho'}\delta_{\alpha\alpha'}}{1-\Sigma_{\alpha\alpha}}.\label{eq:tripcon2}
\end{equation}
Finally, condition (\ref{eq:nambuSC}) leads to
\begin{equation}
[\mathcal{G}_0^{-1}]_{\alpha\alpha}=4-3\Sigma_{\alpha\alpha},\quad [\mathcal{G}_0^{-1}]_{12}=\Sigma_{1 2},\quad L=S.\label{eq:tripcon3}
\end{equation}
Eqs.\,(\ref{eq:tripcon1}), (\ref{eq:tripcon2}), and (\ref{eq:tripcon3}) have to be solved self-consistently for given $A_{\alpha\beta\gamma\delta}$, $B$, $C$ and $D$,
to determine $\mathcal{G}_0^{-1}$ and $L$.

For flavors with piecewise flat profiles (\ref{eq:2flat}), the double-occupancy can be extracted from $Z$ via
\begin{widetext}
\begin{align}
\left<n_{o\uparrow}n_{o\downarrow}\right>&=\frac{1}{4}+\left.\frac{1}{Z}\left(\sum_{\rho\rho'=\pm}\text{or}_{\rho\rho'}\sum_{\alpha\alpha'=1}^2\frac{\partial}{\partial M^{(\uparrow)}_{\alpha\rho,\alpha'\rho'}}\right)\left(\sum_{\rho\rho'=\pm}\text{or}_{\rho\rho'}\sum_{\alpha\alpha'=1}^2\frac{\partial}{\partial M^{(\downarrow)}_{\alpha\rho,\alpha'\rho'}}\right)Z\right|_{M_\sigma=M_o}
\end{align}
\end{widetext}
with
\begin{equation}
\text{or}_{\rho\rho'}=\left\{\begin{array}{ll}-1&\text{if }\rho=\rho'=-1,\\
    1&\text{otherwise.}\end{array}\right.
\end{equation}
\begin{acknowledgments}
This work was supported by the European Research Council (ERC) under Grant Agreement No. 101001310 (SuperProtected). IS thanks the CNRS for financial mobility support.
\end{acknowledgments}

\end{document}